\documentclass[12pt,a4paper,twoside,dvips]{article}
\newcommand{\HOME}{/users/staf/wes/}
\newcommand{\lthreepaper}{\HOME/l3/paper/}
\newcommand{\lthreebiblio}{\lthreepaper biblio/}
\newcommand{\mydirfig}{figs_paper/}
\usepackage{a4p}
\usepackage[usenames]{color}   
\usepackage{cite,mciteplus}
\usepackage{url}
\usepackage{physics}
\usepackage{l3_title}
\usepackage{ifthen}
\usepackage{graphicx}
\usepackage{amssymb}
\usepackage{amsmath}
\usepackage{dcolumn}
\usepackage{pennames}
\usepackage{subscript}
\usepackage{rotating}
\usepackage{longtable}
\usepackage[perpage,symbol*,hang]{footmisc}   
\setfnsymbol{chicago}                         
\makeatletter\def\@makefnmark{\hbox{$^{\@thefnmark}$}}\makeatletter  
\journalname{European Physical Journal C}
\date{May 20, 2011}
\preprint{CERN-PH-EP-2011-080}
\graphicspath{{./}}
\newlength{\capindent}
\newlength{\capwidth}
\newlength{\figwidth}
\newcommand{\icaption}[2][!*!,!]{\hspace*{\capindent}%
  \begin{minipage}{\capwidth}
    \ifthenelse{\equal{#1}{!*!,!}}%
      {\caption{#2}}%
      {\caption[#1]{#2}}
  \end{minipage}}
\newcommand{\adhoc}{\textit{ad hoc}}
 
\newcommand{\PZ}{\ensuremath{\mathrm{Z}}}
 
\newcommand{\JETSET}{{\scshape Jetset}}

\newcommand{\HERWIG}{{\scshape Herwig}}
\newcommand{\PYTHIA}{{\scshape Pythia}}

\newcommand{\ALEPH}{{\scshape aleph}}
\newcommand{\OPAL}{{\scshape opal}}

\newcommand{\Lthree}{{\scshape l}{\small 3}}
\newcommand{\Lthreefoot}{{\scshape l}{\scriptsize 3}}

\newcommand{\BEz}{\text{BE\textsubscript{0}}}
\newcommand{\BEtt}{\text{BE\textsubscript{32}}}

\newcommand{\abs}[1]{\left|#1\right|}
 
\newcommand{\ycut}{\ensuremath{y_\mathrm{cut}}}
\renewcommand{\pt}{\ensuremath{p_\mathrm{t}}}
\newcommand{\mt}{\ensuremath{m_\mathrm{t}}}
\newcommand{\Qsquare}{\ensuremath{Q^2}}

\newcommand{\Qlong}{\ensuremath{Q_\mathrm{L}}}
\newcommand{\Qside}{\ensuremath{Q_\mathrm{side}}}
\newcommand{\Qout}{\ensuremath{Q_\mathrm{out}}}
\newcommand{\qout}{\ensuremath{q_\mathrm{out}}}
\newcommand{\Qle}{\ensuremath{Q_\mathrm{LE}}}

\newcommand{\Qslong}{\ensuremath{Q^2_\mathrm{L}}}
\newcommand{\Qsside}{\ensuremath{Q^2_\mathrm{side}}}
\newcommand{\Qsout}{\ensuremath{Q^2_\mathrm{out}}}
\newcommand{\qsout}{\ensuremath{q^2_\mathrm{out}}}
\newcommand{\Qsle}{\ensuremath{Q^2_\mathrm{LE}}}
\newcommand{\Rsquare}{\ensuremath{R^2}}

\newcommand{\Rlong}{\ensuremath{R_\mathrm{L}}}
\newcommand{\Rside}{\ensuremath{R_\mathrm{side}}}
\newcommand{\Rout}{\ensuremath{R_\mathrm{out}}}

\newcommand{\rhoout}{\ensuremath{\rho_\mathrm{out}}}
\newcommand{\rout}{\ensuremath{r_\mathrm{out}}}

\newcommand{\Rslong}{\ensuremath{R^2_\mathrm{L}}}
\newcommand{\Rsside}{\ensuremath{R^2_\mathrm{side}}}
\newcommand{\Rsout}{\ensuremath{R^2_\mathrm{out}}}
\newcommand{\rhosout}{\ensuremath{\rho^2_\mathrm{out}}}
\newcommand{\rsout}{\ensuremath{r^2_\mathrm{out}}}
\newcommand{\Rle}{\ensuremath{R_\mathrm{LE}}}
\newcommand{\Rsle}{\ensuremath{R^2_\mathrm{LE}}}

\newcommand{\epsillong}{\ensuremath{\epsilon_\mathrm{L}}}
\newcommand{\epsilside}{\ensuremath{\epsilon_\mathrm{side}}}
\newcommand{\epsilout}{\ensuremath{\epsilon_\mathrm{out}}}
\newcommand{\epsille}{\ensuremath{\epsilon_\mathrm{LE}}}

\newcommand{\Deltataulong}{\ensuremath{\Delta\tau_\mathrm{L}}}
\newcommand{\Deltatauside}{\ensuremath{\Delta\tau_\mathrm{side}}}
\newcommand{\Deltatauout}{\ensuremath{\Delta\tau_\mathrm{out}}}
 
\newcommand{\rx}{\ensuremath{r_\mathrm{\kern -0.14em x}}}
\newcommand{\ry}{\ensuremath{r_\mathrm{\kern -0.14em y}}}
\newcommand{\rz}{\ensuremath{r_\mathrm{\kern -0.14em z}}}
\newcommand{\px}{\ensuremath{p_\mathrm{x}}}
\newcommand{\py}{\ensuremath{p_\mathrm{y}}}
\newcommand{\pz}{\ensuremath{p_\mathrm{z}}}
 
\newcommand{\taumodel}{$\tau$-model}
\newcommand{\boldtaumodel}{\boldmath{$\tau$}-model}
 
\newcommand{\chisq}{\ensuremath{\chi^2}}

\newcommand{\Eq}[1]{Eq.\,(\ref{#1})}%
\newcommand{\Eqs}[1]{Eqs.\,(\ref{#1})}%
\newcommand{\Fig}[1]{Fig.\,\ref{#1}}%
\newcommand{\Figs}[1]{Figs.\,\ref{#1}}%
\newcommand{\Tab}[1]{Table~\ref{#1}}%
\newcommand{\Tabs}[1]{Tables~\ref{#1}}%
\newcommand{\dd}[1]{\ensuremath{\,{\mathrm{d}#1}}}%
\newcommand{\alphas}{\ensuremath{\alpha_\mathrm{s}}}

\renewcommand{\ee}{\ensuremath{\mathrm{e^{+}e^{-}}}}

\newcommand{\Evis}{\ensuremath{E_{\mathrm{vis}}}}

\newcommand{\Ra}{\ensuremath{R_\mathrm{a}}}
\newcommand{\invGeV}{\GeV\ensuremath{^{-1}}}
\newcommand{\nbarinv}{\ensuremath{\frac{1}{\bar{n}}}}
 
\newcommand{\pho}{\phantom{0}}
\newcommand{\phm}{\phantom{-}}
\newcommand{\Letter}{article}

\begin{document}
\begin{titlepage}
\title{Test of the \boldmath{$\tau$}-Model
                      of Bose-Einstein Correlations \\
       and Reconstruction of the Source Function     \\
       in Hadronic Z-boson Decay at LEP} %
 
\author{The L3 Collaboration}
%
 
\begin{abstract}
\vspace{-5mm}
Bose-Einstein correlations of pairs of identical charged pions
produced in hadronic Z decays are analyzed in terms of various parametrizations.
A good description is achieved using a L\'evy stable distribution in conjunction with a
model where a particle's momentum is correlated with its space-time point of
production, the \taumodel.
Using this description and the measured rapidity and transverse momentum distributions,
the space-time evolution of particle emission in two-jet events is reconstructed.
However, the elongation of the particle emission region previously observed
is not accommodated in the \taumodel, and
this is investigated using an \adhoc\/ modification. 
\end{abstract}
 
\submitted                   
 
\end{titlepage}

\normalsize
\pagenumbering{roman}  \setcounter{page}{2}
\cleardoublepage
 
\pagenumbering{arabic}
 
\setcounter{page}{1}

\section{Introduction}\label{sect:intr}
 
In particle and nuclear physics, intensity interferometry provides a direct
experimental method for the determination of sizes, shapes and lifetimes
of particle-emitting sources
(for reviews see \cite{Gyulassy:1979,Boal:1990,Baym:1998,Wolfram:Zako2001,Tamas:HIP2002}).
In particular, boson interferometry provides a powerful tool for the
investigation of the space-time structure of particle production processes,
since Bose-Einstein correlations (BEC) of two identical bosons,
first observed in 1959~\cite{Goldhaber:1959,GGLP:1960} and most recently at the Large Hadron
Collider~\cite{CMS:be1,ALICE:be1,CMS:be2,ALICE:be2,ALICE:be3}, reflect both
geometrical and dynamical properties of the particle-radiating source.
 
In \Pep\Pem\ annihilation BEC have been observed\cite{TASSO:1986}
to be maximal when the invariant momentum difference of the bosons,
$Q=\sqrt{-(p_1-p_2)^2}$, is small,
even when one of the relative momentum components is large.
This is not the case either in hadron-hadron interactions \cite{NA22q}
or in heavy-ion interactions \cite{Solz:thesis,Solz:paper},
where BEC are found not to depend simply on $Q$, but to
decrease when any of the relative momentum components is large,
a behavior that can be described by hydrodynamical models of the source
\cite{Tamas;Lorstad:1996,Tamas:HIP2002}.
 
The size (radius) of the source in heavy-ion collisions
has been found to decrease with increasing transverse momentum, \pt,
or transverse mass, $\mt=\sqrt{m^2+\pt^2}$, 
of the bosons.  This effect can also be explained by hydrodynamical
models\cite{Tamas;Lorstad:1996,Wiedemann:96}.
A similar effect has been seen in pp collisions\cite{STAR:QM05}, as well as
in \Pep\Pem\ annihilation~\cite{Smirnova:Nijm96,Dalen:Maha98,OPAL:2007}.

A model for BEC in \Pep\Pem\ annihilation, known as the \taumodel,
which results in BEC depending simply on $Q$, rather than on its components separately,
has been proposed\cite{Tamas;Zimanji:1990}.
In this model the simple $Q$ dependence is a consequence of a strong correlation
between a particle's four-momentum and its space-time point of production.
This strong correlation also leads to a decrease of the observed source size
with increasing \mt\cite{ourTauModel}.
 
On the other hand, studies of BEC in \Pep\Pem\ annihilation at LEP have found that the BEC
correlation function does depend on components of $Q$, the shape of the source being elongated along the event
axis~\cite{L3_3D:1999,OPAL3D:2000,DELPHI2D:2000,ALEPH:2004,OPAL:2007}.  At HERA a similar elongation
is observed in neutral current deep inelastic ep scattering~\cite{ZEUS2D:2004}.
Note that here ``source'' refers not to the entire volume in which particles are emitted, but rather the
smaller ``region of homogeneity'' 
from which pions are emitted that have momenta similar enough to interfere and contribute to the correlation function.
The size of this region of homogeneity is sometimes referred to as the correlation length.
 
The main purpose of the present paper is to test the \taumodel, but the question of to what extent BEC in
\Pep\Pem\ annihilation depend differently on different components of $Q$ is also considered.
%
Here we study BEC in hadronic \PZ\ decay.
We investigate various static parametrizations in terms of $Q$.
Fitting over a larger $Q$ range than in previous studies, we find that none of these parametrizations
gives an adequate description of the Bose-Einstein correlation function.
Next we investigate the \taumodel\ and find that it provides a good description of BEC,
both for two-jet and three-jet events.
The results for two-jet events,
from fits in terms of $Q$ and the transverse masses of the two pions with respect to the event axis,
are used to reconstruct the complete
space-time picture of the particle emitting source of two-jet events in hadronic Z decay.
Finally, the relevance of the elongation mentioned above is investigated using an \adhoc\/ modification of the \taumodel.

\section{Analysis}                     \label{sect:anal}
\subsection{Data} \label{sect:data}
The data used in the analysis were collected by the
\Lthree\ detector~\cite{l3:construction,L3:Ecal_calib,L3:Hcal,L3:TEC,l3:SMD}
at an \Pep\Pem\    center-of-mass energy,  $\sqrt{s}$, of about 91.2\,\GeV.
Calorimeter clusters having energy greater than 0.1\,\GeV\ are used to determine the event thrust\cite{thrustdef}
axis and to
classify events as two- or three-jet events, which are     analyzed separately,
since differences in the space-time structure  (and hence in the BEC parameters)
could be expected, and indeed have been observed \cite{OPALmult:1996,OPAL3D:2000}.
 
Backgrounds to hadronic \PZ\ decays
such as leptonic \PZ\ decays, beam-wall and beam-gas interactions, and two-photon events
are excluded by requiring that the visible energy, \Evis,  be within $0.5\sqrt{s}$ and
$1.5\sqrt{s}$,
that the transverse and longitudinal energy imbalances be less than $0.6\Evis$ and $0.4\Evis$, respectively,
and that the number of calorimeter clusters be at least 15.
In order to ensure well-measured charged tracks, events are required to have the thrust
direction within the barrel region of the calorimeters and the acceptance region of the tracking chamber:
$\abs{\cos(\Theta)}<0.74$, where $\Theta$ is the angle between the thrust axis and the beam
direction. High precision charged tracks are selected by requiring
\begin{itemize}
  \item transverse momentum greater than 150 \MeV,
  \item at least one hit in the inner region of the tracking chamber (TEC),
  \item more than 25 hits in the entire TEC spanning at least 40 wires of the possible 62,
  \item a distance, in the plane transverse to the beam, of closest approach to the interaction vertex
        less than 10 mm.
\end{itemize}
Further, tracks lying in two small regions of     azimuthal angle having less precise calibration are
rejected: $45^\circ$--$52.5^\circ$ and $225^\circ$--$232.5^\circ$.
Since the resolution of the opening angle between pairs of tracks is crucial for the study of BEC,
tracks are also rejected if there is no hit in the Z-chamber of the TEC.
To further reject $\tau^+\tau^-$ events, the second largest angle, $\phi_2$, between any
two neighboring tracks in the transverse plane is required to be in the range $20^\circ$--$170^\circ$.
To further ensure that the event is well contained within the acceptance of the TEC, the thrust axis is
determined using only the high precision tracks, and the event is rejected if
$\abs{\cos(\Theta_\mathrm{TEC})}>0.7$, where $\Theta_\mathrm{TEC}$ is the angle between this thrust axis
and the beam.
 
In total about 0.8 million events with an average number of about 12 high precision charged tracks are
selected. This results in approximately 36 million like-sign pairs of charged tracks.
 
The number of jets in an event is determined using the
Durham jet algorithm~\cite{durham,durham2,durham3}    
with a jet resolution parameter $\ycut=0.006$, yielding about 0.5 million two-jet events and 0.3 million events
having more than two jets.
Since there are few events with more than three jets, the category of events with more than two jets is
referred to as the three-jet sample.
 
\subsection{Bose-Einstein Correlation Function}
The two-particle correlation function of two particles with
four-momenta $p_{1}$ and $p_{2}$ is given by the ratio of the two-particle number density,
$\rho_2(p_{1},p_{2})$,
to the product of the two single-particle number densities, $\rho_1 (p_{1})\rho_1 (p_{2})$.
Since we are here interested only in the correlation $R_2$ due to Bose-Einstein
interference, the product of single-particle densities is replaced by
$\rho_0(p_1,p_2)$,
the two-particle density that would occur in the absence of Bose-Einstein correlations:
\begin{equation} \label{eq:R2def}
  R_2(p_1,p_2)=\frac{\rho_2(p_1,p_2)}{\rho_0(p_1,p_2)} \;.
\end{equation}
This $\rho_2$ is corrected for detector acceptance and efficiency using Monte Carlo events
generated by the \JETSET\  Monte Carlo generator~\cite{JETSET74}
with the so-called \BEz\ simulation of BEC~\cite{BE0}
to which a full detector simulation has been applied \cite{detsim}, by multiplying the measured $\rho_2$, on
a bin-by-bin basis, by the ratio of $\rho_2$ of the generated events to
$\rho_2$ of the generated events after detector simulation.
For the detector-simulated events, as for the data, all charged tracks are used,
whereas for the generator-level events only charged pions are used, since all measured tracks are assumed to
be pions in the calculation of $Q$.
Thus the Monte Carlo generator is used to extract the $Q$ distribution for equally charged pion pairs from the
observed $Q$ distribution of all equally charged particle pairs.

An event mixing technique is used to construct $\rho_0$, whereby all tracks of each data
event are replaced by  tracks from  different events having a multiplicity similar to that of the original event.
This is accomplished by first rotating each event to a frame whose axes are the thrust, major, and minor~\cite{majorminor}
directions and then assigning the events to classes based on the track multiplicity.  Each multiplicity
defines a class except for very low and very high multiplicities, in which case several adjacent
multiplicities are grouped into the same class in order to obtain sufficient statistics.
The replacement track is randomly chosen from tracks of the same sign in a randomly chosen event either in
the same class as the data event or a neighboring or next-to-neighboring class.
Ideally one would define the classes based on the total particle (charged plus neutral) multiplicity, since
the aim is to use events where the available phase space of the particles is the same.
However, the ratio of neutral to charged multiplicities is not constant and both multiplicities are subject to detector efficiency losses.
In an attempt to take this into account we therefore also use events of nearby classes.

A correction for detector acceptance and efficiency is applied to $\rho_0$ in the same way as to $\rho_2$.
The mixing technique removes all correlations, \eg, resonances and energy-momentum conservation,
not just Bose-Einstein correlations.
Hence, $\rho_0$  is also corrected for this  
by a multiplicative factor which is the ratio of the densities of events to mixed events found using
events generated by \JETSET\ without BEC simulation.
 
Including all corrections, $R_2$ is measured by
\begin{equation} \label{eq:R2cor}
  R_2 = \left(R_\textrm{2\;data} R_\textrm{2\;gen}\right) / \left(R_\textrm{2\;det} R_\textrm{2\;gen-noBE}\right) \;,
\end{equation}
where data, gen, det, gen-noBE refer, respectively, to the data sample, a generator-level Monte Carlo sample,
the same Monte Carlo sample passed through detector simulation and subjected to the same selection procedure as the data,
and a generator-level sample of a Monte Carlo generated without BEC simulation.
 
We note that $R_2$ is unaffected by single-particle acceptances and efficiencies. Being the same for $\rho_2$
and $\rho_0$, these cancel in the ratio.
Systematic uncertainties on $R_2$ are highly correlated point-to-point.
Hence, in performing fits to $R_2$, only statistical uncertainties
will be used in calculating the \chisq\ which is minimized.
 
In principle $R_2$ should also be corrected for final-state interactions (FSI), both Coulomb and strong, or
alternatively, the functional form used to fit $R_2(Q)$ should be adapted to take account of
FSI\cite{Bowler:1988,Osada:1996}.
Coulomb interactions, being repulsive for like-sign charged pions, serve to increase the measured values of $Q$.
This is often taken into account by weighting the pion pairs with the inverse of the so-called Gamow
factor\cite{Gyulassy:1979}.
However, this factor strongly over-compensates, since it is derived assuming that the source is point-like
and that all particles are pions\cite{Bowler:1991}.
Further the strong \Pgp-\Pgp\ FSI at short distances are attractive, further reducing
the effect\cite{Bowler:Marburg}.  The net effect is small, amounting to about 3\% at $Q=0$\cite{Osada:1996},
comparable to the statistical uncertainty in the first bin of $R_2(Q)$ (\cf\ \Figs{fig:gauss_2jet}--\ref{fig:gauss_3jet}).
Further, if $R_\textrm{2\;data}$ is weighted to account for FSI,
  $R_\textrm{2\;gen}$, $R_\textrm{2\;det}$ and $R_\textrm{2\;gen-noBE}$
need to be
similarly weighted.
This is because
the Monte Carlo programs
are tuned to data, and  this tuning presumably accounts, at least partially, for the effects
of FSI, even though the Monte Carlo program does not explicitly do so.
The result is that the various weights, to a large degree, cancel.
We have used unlike-sign charged pions to check that applying Gamow factors is not needed in the present analysis.
%
The uncertainty on FSI is included in the systematic uncertaities by varying the fit range (\cf\ Section \ref{sect:syst}).
 
\section{Parametrizations of BEC}              \label{sect:param}
 
\subsection{Dependence on \boldmath{$Q$}} \label{sect:Q}
Since the four-momenta of the two particles are on the mass-shell,
the correlation function is defined in six-dimensional momentum space.
Since BEC   can be large only at small four-momentum difference
$Q$,
they are often parametrized in this one-dimensional distance measure.
With a few assumptions~\cite{GGLP:1960,Boal:1990,Tamas:HIP2002},
the two-particle correlation function, \Eq{eq:R2def},
is related to the Fourier transformed source distribution:
\begin{equation}  \label{eq:R2fourier}
     R_2(Q)       = \gamma \left[ 1 + \lambda \abs{\tilde{f} (Q)}^2 \right]
                    \left(1 + \epsilon Q \right) \;,
\end{equation}
where $f(x)$ is the (space-time) density distribution of the source,
 and
 $\tilde{f}(Q)$ is the Fourier transform of $f(x)$.
The parameter $\lambda$ is introduced as a measure of the strength of the correlation.
It may be less than unity for a variety of reasons such as inclusion of misidentified non-identical particles
or the presence of long-lived resonance decays if the particle emission
consists of a small, resolvable core and a halo having experimentally unresolvable large length scales
\cite{Bolz:1992hc,Csorgo:1994in}.
The parameter $\gamma$ and the $(1 + \epsilon Q)$ term parametrize possible long-range correlations not
adequately accounted for in the reference sample.  While there is no guarantee that $(1 + \epsilon Q)$
is the correct form, we will see that it does provide a good description of $R_2$ in the region $Q>1.5\GeV$.
In fact, if the reference sample indeed removes only BEC, we expect $R_2$ to contain no long-range correlations
and hence to find $\epsilon=0$.
 
There is no reason, however,
to expect the hadron source to be spherically symmetric in jet fragmentation.
Recent investigations have, in fact, found an elongation of the source along the
jet axis~\cite{L3_3D:1999,OPAL3D:2000,DELPHI2D:2000,ALEPH:2004,OPAL:2007}.
While this effect is      well established, the elongation is actually only about 20\%,
which suggests that a parametrization in terms of the single variable $Q$,
may be a good approximation.
 
\begin{figure}
  \centering
  \includegraphics[width=.5\figwidth]{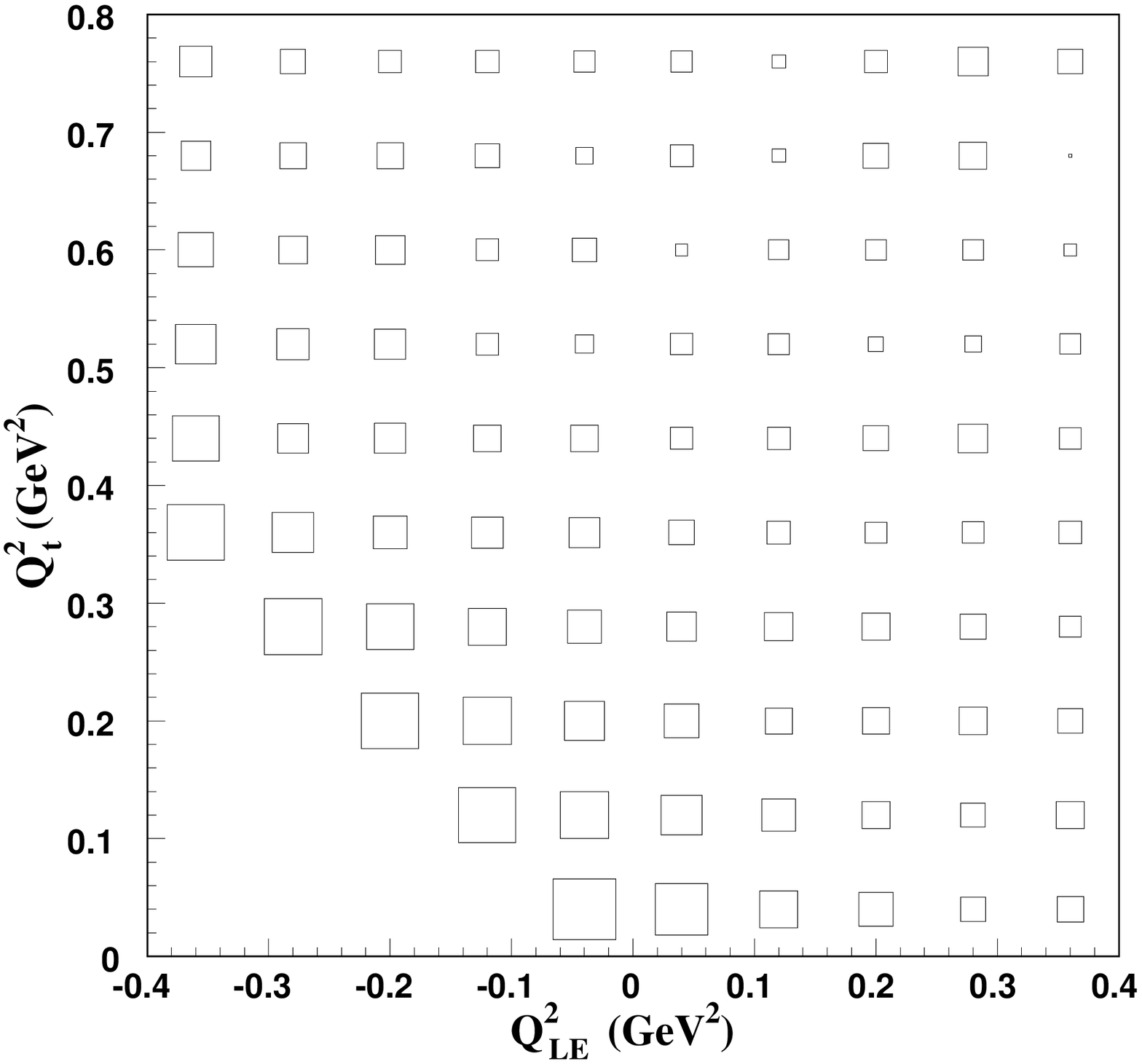}
  \includegraphics[width=.5\figwidth]{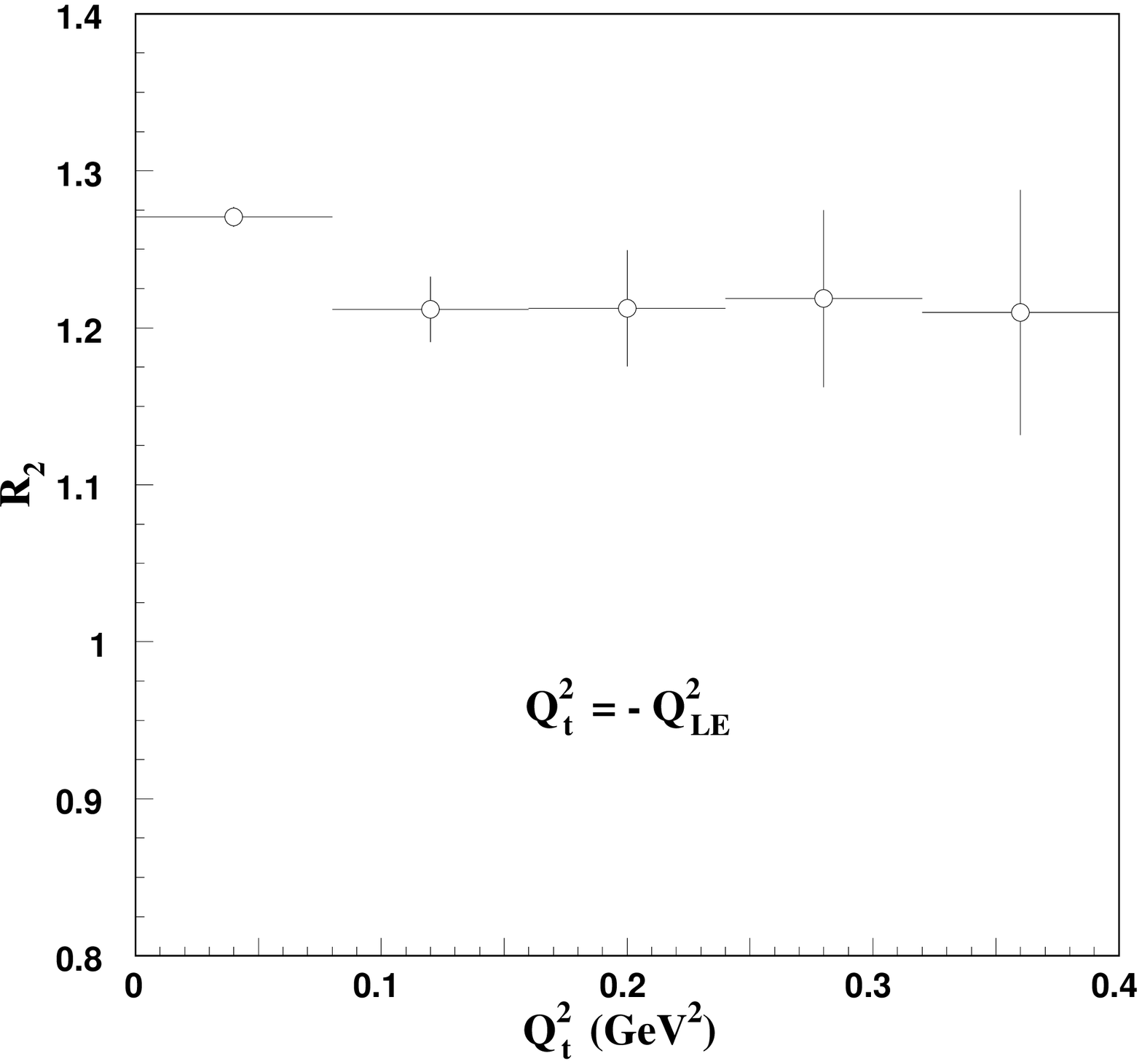}
   \caption{(a) $R_2$ for two-jet events as function of the squares of the transverse momentum difference,
              $Q_\mathrm{t}^2=(\vec{p}_\mathrm{t1}-\vec{p}_\mathrm{t2})^2$,
                and the combination of longitudinal momentum difference and energy difference,
              $\Qsle=(p_\mathrm{l1}-p_\mathrm{l2})^2-(E_1-E_2)^2$.
            (b)  $R_2$ \vs\ $Q_\mathrm{t}^2$ when $Q_\mathrm{t}^2\approx -\Qsle$,
                which corresponds to $Q^2\approx0$.
           \label{fig:qinv}
           }
\end{figure}

This is not the case in heavy-ion and hadron-hadron interactions, where BEC are found not to depend simply
on $Q$, but on components of the four-momentum difference separately
\cite{Tamas:HIP2002, Tamas;Lorstad:1996, NA22q,Solz:thesis,Solz:paper,STAR:QM05}.
However, in \Pep\Pem\ annihilation at lower energy~\cite{TASSO:1986} it has been observed that $Q$ is the
appropriate variable.
We checked this \cite{tamas:thesis} both for all and for two-jet events:
We observe that $R_2$ does not decrease when both $q^2=(\vec{p}_1-\vec{p}_2)^2$ and
$q_0^2=(E_1-E_2)^2$ are large while $Q^2=q^2-q_0^2$ is small, but is maximal for
$Q^2=q^2-q_0^2=0$,
independent of the individual values of $q$ and $q_0$.
In a different decomposition, $Q^2=Q_\mathrm{t}^2 + \Qsle$, where
$Q_\mathrm{t}^2=(\vec{p}_\mathrm{t1}-\vec{p}_\mathrm{t2})^2$ is the component transverse to the thrust axis
and
$\Qsle=(p_\mathrm{l1}-p_\mathrm{l2})^2-(E_1-E_2)^2$ combines the longitudinal momentum and
energy
differences, we find
$R_2$ to be maximal along the line $Q=0$, as is shown in \Fig{fig:qinv},
and fits, though of poor \chisq\ (confidence level of about 1\%), are consistent with equal radii,
as has previously been observed by \ALEPH~\cite{ALEPH:1992}.
 
We conclude that a parametrization in terms of  $Q$
can be considered a reasonable approximation for the purposes of this \Letter.
We shall return to the question of elongation in Section~\ref{sect:elongation}.  

\subsection{Symmetric parametrizations} \label{sect:sym_param}
The simplest assumption is that the source 
has a symmetric Gaussian distribution with mean $\mu=0$ and standard deviation $R$.
In this case
$\tilde{f} (Q)=\exp \left(i\mu Q -\frac{(RQ)^2}{2}\right)$ and
\begin{equation} \label{eq:gaussR2}
  R_2(Q) = \gamma \left[1 + \lambda \exp \left(-(R Q)^2 \right) \right]
           \left(1 + \epsilon Q \right) \;.
\end{equation}
However, this parametrization is often found to be inadequate in its description of data.
 
A model-independent way to study deviations from the Gaussian parametrization is to
use~\cite{Tamas:Moriond28,Tamas:Cracow1994,Tamas:HIP2002}
the Edgeworth expansion~\cite{Edgeworth} about a Gaussian. 
Keeping only the first non-Gaussian term, we have
\begin{equation} \label{eq:edgeworthR2}
  R_2(Q) = \gamma \left(1 + \lambda \exp \left(-(R Q)^2 \right)
                  \left[1+\frac{\kappa}{3!}H_{3}(RQ)\right]
                  \right)
           \left(1 + \epsilon Q \right) \;,
\end{equation}
where
$\kappa$ is the third-order cumulant moment and
$H_{3}(RQ)\equiv (\sqrt{2}RQ)^{3}-3\sqrt{2}RQ$
is the third-order Hermite polynomial.
Note that the second-order cumulant corresponds to the radius~$R$.
\Eq{eq:edgeworthR2} reduces to \Eq{eq:gaussR2} for $\kappa=0$.
 
Another way to depart from the assumption of a Gaussian source is to replace the Gaussian by a
symmetric L\'evy stable distribution, which
is characterized by three parameters: $x_0$, $R$, and $\alpha$.
Its Fourier transform,
$\tilde{f}(Q)$, has the following form:
\begin{equation}
  \tilde{f}(Q) = \exp \left(iQ x_0 - \frac{|R Q|^\alpha}{2} \right) \;,
\end{equation}
where the index of stability, $\alpha$, satisfies the inequality $0<\alpha \leq 2$.
The case $\alpha=2$ corresponds to a Gaussian source distribution with mean $x_0$ and standard deviation $R$.
For more details, see, \eg, \cite{Nolan:2010}.
Then $R_2$ has the following form~\cite{Tamas:Levy2004}:
\begin{equation}\label{eq:symlevR2}
    R_2(Q) = \gamma \left[ 1+ \lambda \exp \left(-(RQ)^\alpha \right) \right]
             (1+ \epsilon Q) \;.
\end{equation}
 
\begin{sidewaysfigure}
  \centering
  \includegraphics[width=.32\textwidth,bb=58 87 519 682,clip]{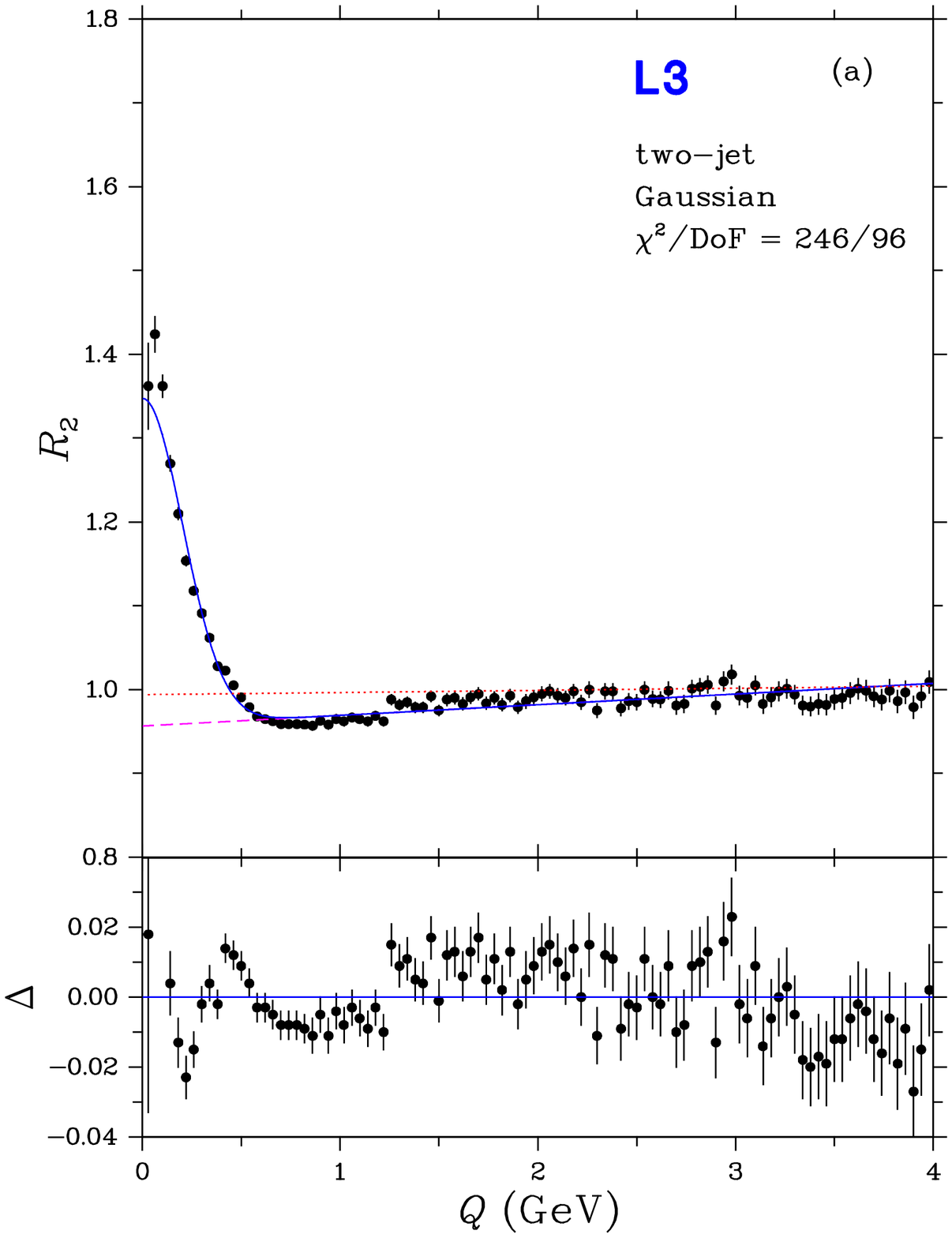}
  \includegraphics[width=.32\textwidth,bb=58 87 519 682,clip]{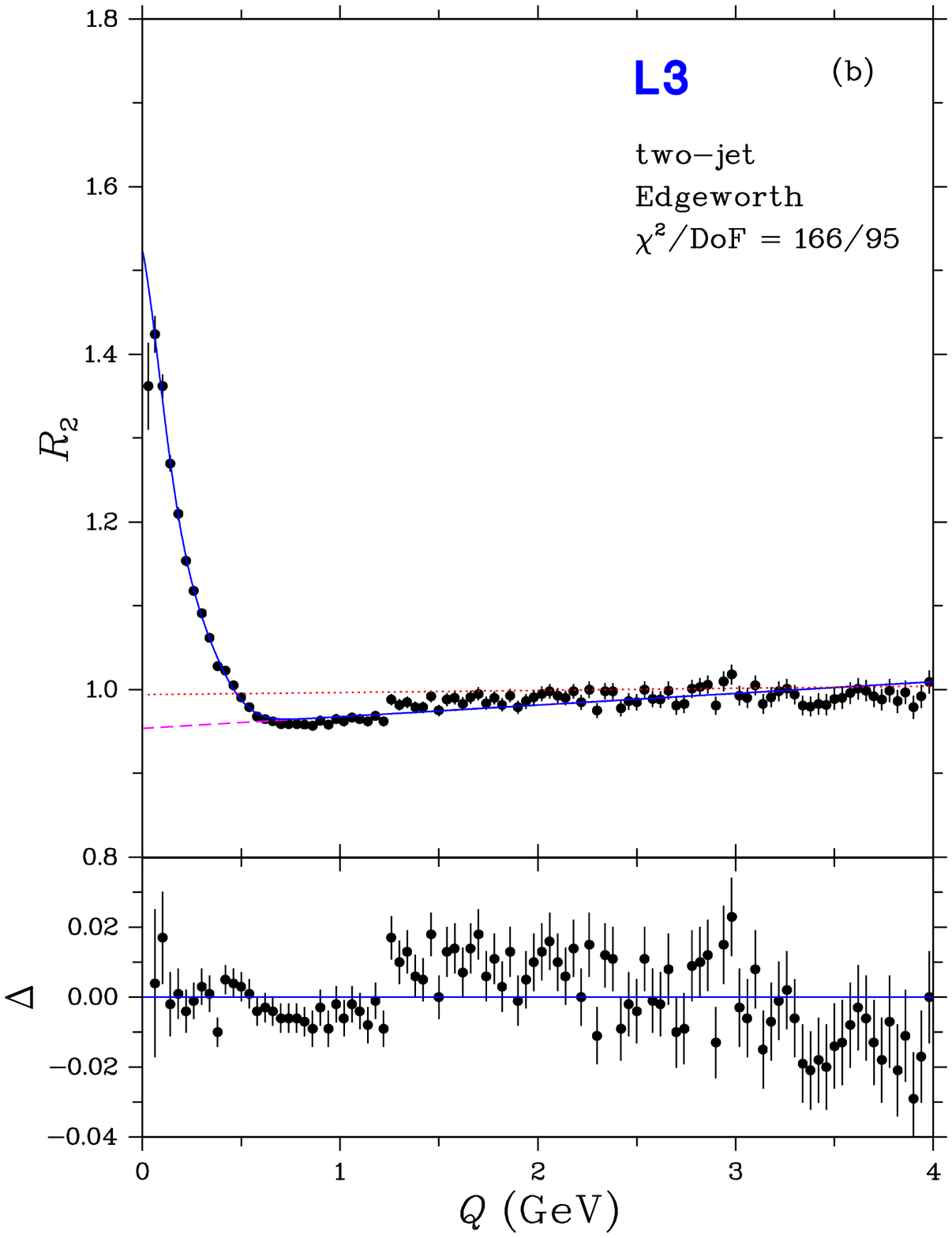}
  \includegraphics[width=.32\textwidth,bb=58 87 519 682,clip]{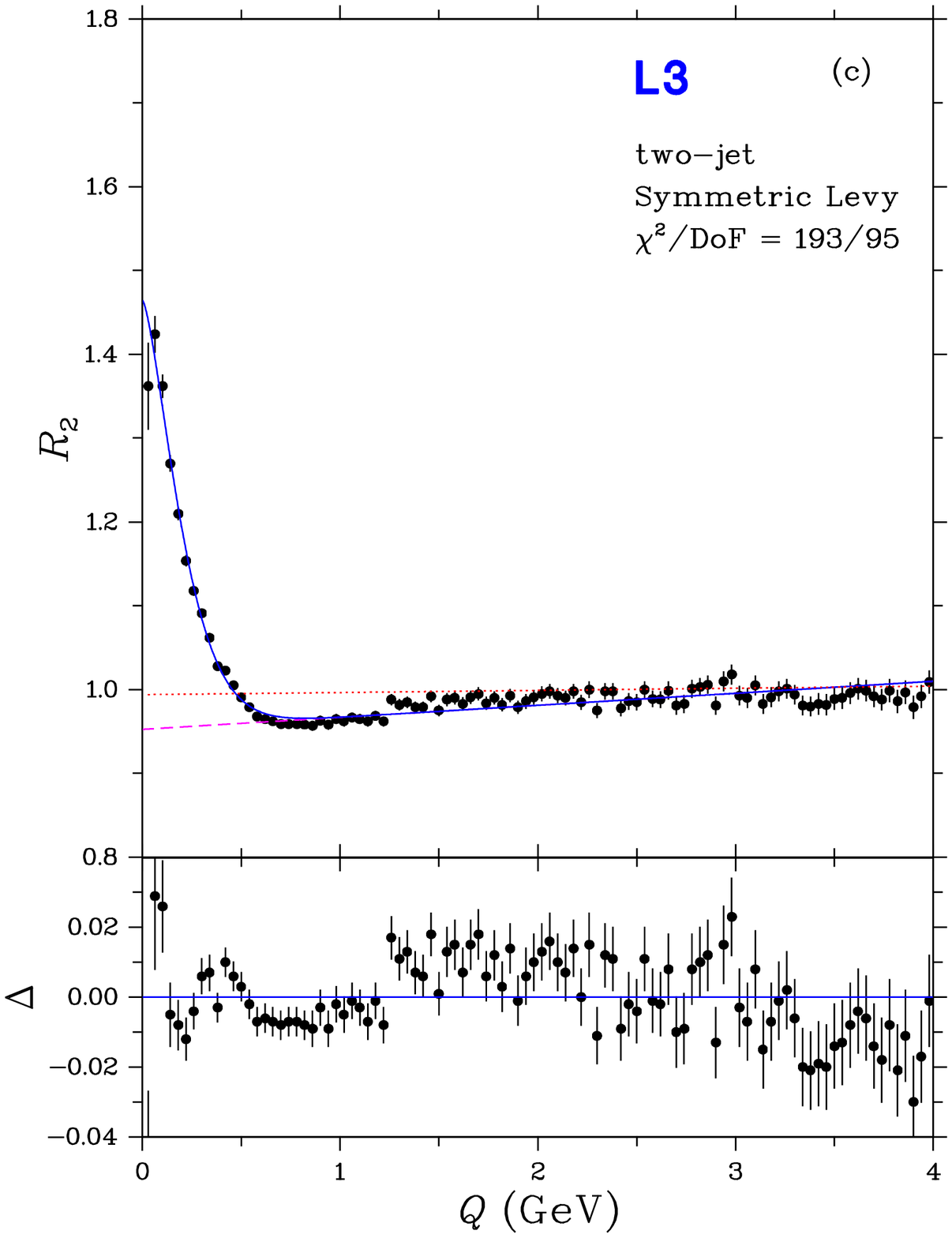}
  \caption{The Bose-Einstein correlation function $R_2$ for two-jet events with the
results of fits of
           (a) the Gaussian, \Eq{eq:gaussR2},
           (b) the Edgeworth, \Eq{eq:edgeworthR2}, and
           (c) the symmetric L\'evy, \Eq{eq:symlevR2}, parametrizations.
           Also plotted is $\Delta$, the difference between the fit and the data.
           The dashed line represents the long-range part of the fit, \ie, $\gamma(1+\epsilon Q)$.
           The dotted line represents a fit with $\lambda=0$ to $Q>1.52\;\GeV$.
           \label{fig:gauss_2jet}
           }
\end{sidewaysfigure}

\begin{sidewaysfigure}
  \centering
  \includegraphics[width=.32\textwidth,bb=58 87 519 682,clip]{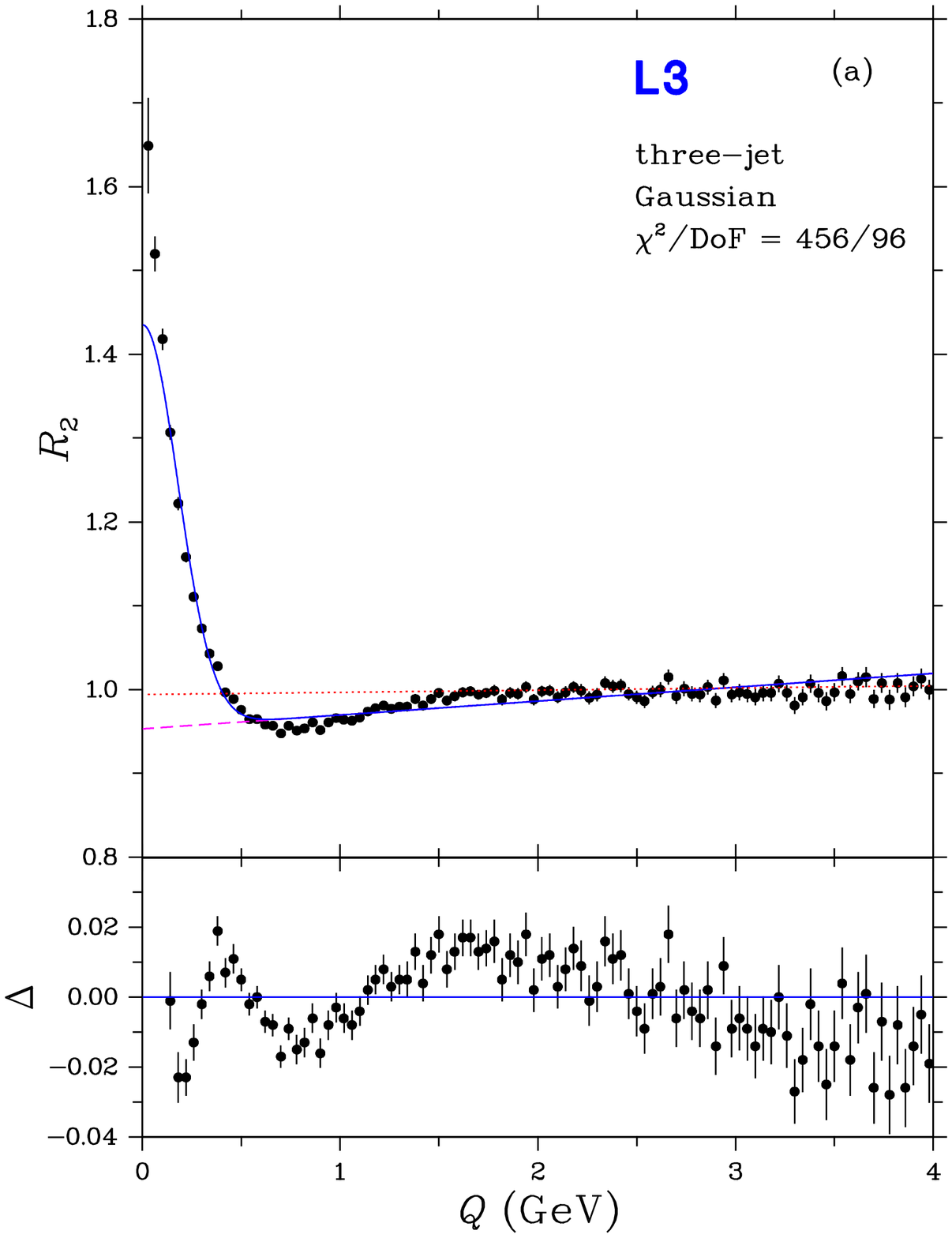}
  \includegraphics[width=.32\textwidth,bb=58 87 519 682,clip]{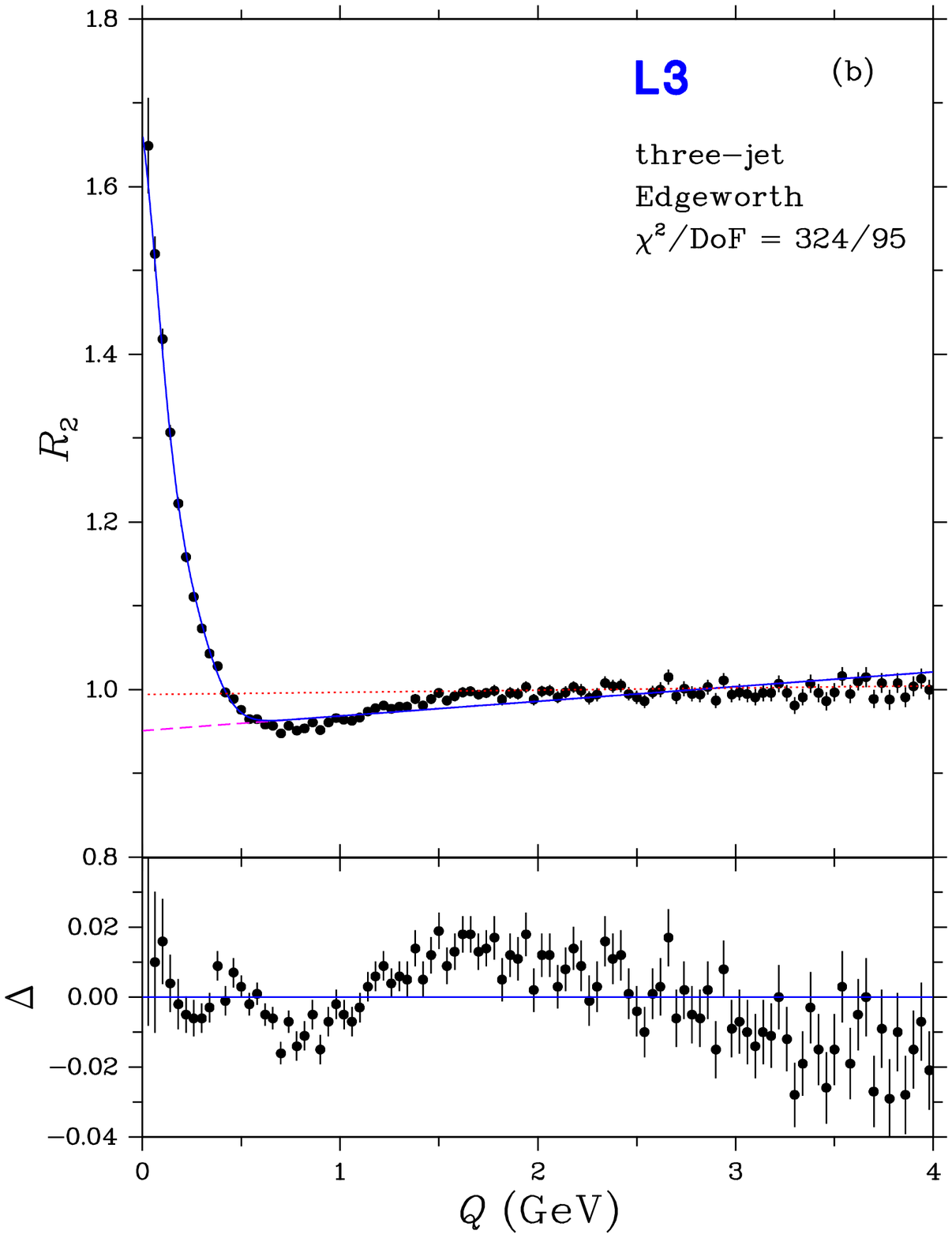}
  \includegraphics[width=.32\textwidth,bb=58 87 519 682,clip]{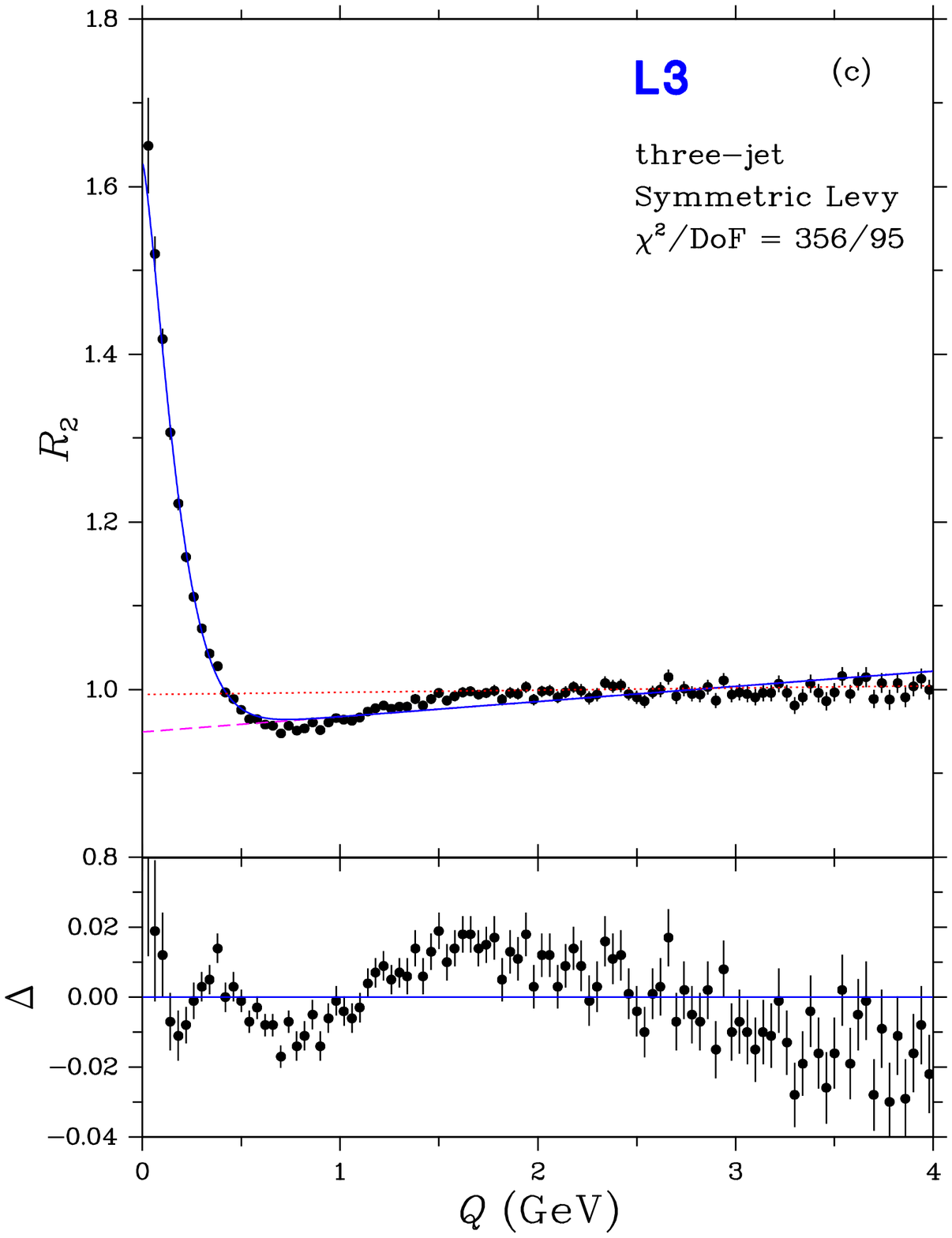}
  \caption{The Bose-Einstein correlation function $R_2$ for three-jet events with the
           results of fits of
           (a) the Gaussian, \Eq{eq:gaussR2},
           (b) the Edgeworth, \Eq{eq:edgeworthR2}, and
           (c) the symmetric L\'evy, \Eq{eq:symlevR2}, parametrizations.
           Also plotted is $\Delta$, the difference between the fit and the data.
           The dashed line represents the long-range part of the fit, \ie, $\gamma(1+\epsilon Q)$.
           The dotted line represents a fit with $\lambda=0$ to $Q>1.52\;\GeV$.
           \label{fig:gauss_3jet}
           }
\end{sidewaysfigure}

These three parametrizations are fitted to the data for both two- and three-jet events.
Fits of the Gaussian parametrization, \Eq{eq:gaussR2}, to the data (shown in \Figs{fig:gauss_2jet}a
and~\ref{fig:gauss_3jet}a for two- and three-jet events, respectively)
result in unacceptably low confidence levels: $10^{-15}$ ($\chisq=246$ for 96 degrees of freedom) for two-jet
 and much worse ($\chisq=456$) for three-jet events.
The fit is particularly bad at low values of $Q$,
where both for two- and three-jet events $R_2$ is much steeper than a Gaussian.

A fit of the Edgeworth parametrization, \Eq{eq:edgeworthR2}, to the two-jet data, shown in \Fig{fig:gauss_2jet}b,
finds $\kappa=0.74\pm0.07$, about 10 standard deviations from the Gaussian value of zero.
The confidence level is indeed
much better than that of the purely Gaussian fit, but is still poor, approximately $10^{-5}$.
Close inspection of the figure shows that the fit curve is systematically above the data in the region
0.6--1.2\,\GeV\ and that the data for $Q\ge1.5$\,\GeV\ appear flatter than the curve, as is also the case
for the purely Gaussian fit.
Similar behavior, but with a worse \chisq, is observed for three-jet events
(\Fig{fig:gauss_3jet}b).

The fit of the L\'evy parametrization, \Eq{eq:symlevR2}, to the two-jet data, shown in \Fig{fig:gauss_2jet}c,
finds $\alpha = 1.44\pm0.06$, far from the Gaussian value of 2.
The confidence level, although
improved compared to the fit of \Eq{eq:gaussR2}
is still unacceptably low, approximately $10^{-8}$.
Similar behavior, with a worse \chisq, is seen for three-jet events (\Fig{fig:gauss_3jet}c).

Both the symmetric L\'evy parametrization and the Edgeworth parametrization  do a fair job of describing the
region $Q<0.6\,\GeV$, but fail at higher $Q$.  $R_2$ in the region $Q\ge1.5\GeV$ is nearly constant
($\approx1$). However, in the region 0.6--1.5\,\GeV\ $R_2$ has a smaller value,
dipping below unity\footnote{More correctly, dipping below the value of $\gamma(1+\epsilon)$.},
which is indicative of an anti-correlation.
This is clearly seen
in \Figs{fig:gauss_2jet} and~\ref{fig:gauss_3jet}
by comparing the data in this region  to an extrapolation of a linear fit, \Eq{eq:gaussR2} with
$\lambda=0$, in the region $Q\ge1.5\,\GeV$.
The inability to describe this dip in $R_2$ is the primary reason for the failure of both the
Edgeworth and symmetric L\'evy parametrizations.
Failure to describe this anti-correlation region, \ie, the dip below unity, is an inherent feature of
any parametrization of the form \Eq{eq:R2fourier}.
We note that this dip is less apparent if one only plots (and fits) $R_2$ for $Q<2\,\GeV$ as has usually been
done in the past, the dip being accommodated by a relatively large value of $\epsilon$.

\subsection{Time dependence of the source}
The parametrizations discussed so far, which have proved insufficient to describe the BEC, all assume a static source.
The parameter $R$ 
is a constant.
It has, however, been observed that $R$ depends on the transverse mass,
$\mt=\sqrt{m^2+\pt^2}=\sqrt{E^2-\pz^2}$, of the pions~\cite{Smirnova:Nijm96,Dalen:Maha98,OPAL:2007}.
It has been shown~\cite{Bialas:1999,Bialas:2000} that this dependence can be understood if the produced
pions satisfy, approximately, the (generalized) Bjorken-Gottfried
condition~\cite{Gottfried:1972,Bjorken:SLAC73,Bjorken:1973,Gottfried:1974,Low:1978,Bjorken:ISMD94}, whereby
the four-momentum of a produced particle and the space-time position at which it is produced are linearly
related.
%
Such a correlation between space-time and momentum-energy is  also a feature of the Lund string model,
which, as incorporated in \JETSET,
is very successful in describing detailed features of the hadronic final states of \Pep\Pem\ annihilation.
Not only inclusive single-particle distributions and event shapes, but also correlations\cite{L3:angmult,L3:Hq,L3:BP}
are well described.

\subsubsection{The \boldtaumodel}  \label{sect:taumodel}
In section~\ref{sect:Q} we have seen that BEC depend, at least approximately, only on $Q$ and not on
its components separately.
This is a non-trivial result.
For a hydrodynamical type of source,
on the contrary, BEC decrease when any of the relative momentum components is large~\cite{Tamas:HIP2002, Tamas;Lorstad:1996}.
Further, we have seen that $R_2$ shows anti-correlations in the region 0.6--1.5\,\GeV, dipping below its values at higher $Q$.
 
A model which predicts such $Q$-dependence, as well as the absence of dependence on the components of $Q$ separately,
is the so-called \taumodel~\cite{Tamas;Zimanji:1990}.
Further it incorporates the  Bjorken-Gottfried condition and
predicts a specific transverse mass dependence of $R_2$,
which we subject to an experimental test here.
 
In this model, it is assumed that the average production point in the overall center-of-mass system,
$\overline{x}=(\overline{t},\overline{r}_\mathrm{x},\overline{r}_\mathrm{y},\overline{r}_\mathrm{z})$,
of particles with a given four-momentum $p=(E,\px,\py,\pz)$ is given by
\begin{equation} \label{eq:tau-corr}
   \overline{x}^\mu (p^\mu)  = a\tau p^\mu \;.
\end{equation}
In the case of two-jet events,
  $a=1/\mt$
where
\mt\ is the transverse mass
and
$\tau = \sqrt{\overline{t}^2 - \overline{r}_{\kern -0.14em \mathrm{z}}^2}$ is the longitudinal proper time.\footnote{The
terminology `longitudinal' proper time and `transverse' mass seems customary in the literature
even though their definitions are analogous $\tau = \sqrt{\overline{t}^2 - \overline{r}_{\kern -0.14em \mathrm{z}}^2}$
and                                         $ \mt = \sqrt{E^2            - \pz^2}$.}
For isotropically distributed particle production, the transverse mass is replaced by the
mass in the definition of $a$ and $\tau$ by the proper time,
$\sqrt{\overline{t}^2 - \overline{r}_{\kern -0.14em \mathrm{x}}^2
                      - \overline{r}_{\kern -0.14em \mathrm{y}}^2
                      - \overline{r}_{\kern -0.14em \mathrm{z}}^2}$.
In the case of three-jet events the relation is more complicated.
 
The second assumption is that the distribution of $x^\mu (p^\mu)$ about its average,
$\delta_\Delta ( x^\mu(p^\mu) - \overline{x}^\mu (p^\mu) )$, is narrower than the
proper-time distribution, $H(\tau)$.
Then   the two-particle Bose-Einstein correlation function is indeed found to depend on the invariant
relative momentum $Q$, rather than on its separate components, as well as on the values of $a$ of the two particles\cite{ourTauModel}:
\begin{equation} \label{eq:levyR2}
   R_2(p_1,p_2) = 1 +         \mathrm{Re} \widetilde{H}\left(\frac{a_1 Q^2}{2}\right)
                                          \widetilde{H}\left(\frac{a_2 Q^2}{2}\right) \;,
\end{equation}
where $\widetilde{H}(\omega) = \int \mathrm{d} \tau H(\tau) \exp(i \omega \tau)$
is the Fourier transform (characteristic function) of $H(\tau)$.
Note that $H(\tau)$ is normalized to unity.

Since there is no particle production before the onset of the collision,
$H(\tau)$ should be a  one-sided distribution.
In the leading log approximation of QCD  the parton shower
is a fractal~\cite{Dahlqvist:1989yc,Gustafson:1990qi,Gustafson:1991ru}.
Further,  a L\'evy distribution arises naturally from a fractal~\cite{metzler}.
We are thus led to
   choose a one-sided L\'evy distribution for $H(\tau)$~\cite{ourTauModel}.
The characteristic function of $H(\tau)$ can then be written~\cite{Tamas:Levy2004}  (for $\alpha\ne1$)
as\footnote{Our notation, $H(\omega)$, corresponds to Nolan's \cite{Nolan:2010}
$S(\alpha,\beta,\gamma,\delta;1)$ parametrization with $\beta=1$,
$\gamma\Longrightarrow\Delta\tau/2^{1/\alpha}$, and
$\delta\Longrightarrow\tau_0$.
The special case of $\alpha=1$ is also given by Nolan \cite{Nolan:2010}.
}
\begin{equation} \label{eq:levy1sidecharf}
   \widetilde{H}(\omega) = \exp\left[ -\frac{1}{2}\left(\Delta\tau|\omega|\strut\right)^\alpha
          \left( 1 -  i\, \mathrm{sign}(\omega) \tan\left(\frac{\alpha\pi}{2}\right) \strut \right)
       + i\,\omega\tau_0\right]
 \; ,
\end{equation}
where the parameter $\tau_0$ is the proper time of the onset of particle production
and $\Delta \tau$ is a measure of the width of the proper-time distribution.
Using this characteristic function in \Eq{eq:levyR2},
and incorporating the factor $\lambda$ and the long-range parametrization,
yields
\begin{equation} \label{eq:levyR2a}
\begin{split}
   R_2(Q,a_1,a_2) &= {} \gamma \left\{  1 +
      \lambda\cos\left[\frac{\tau_0 Q^2 (a_1+a_2)}{2} +
\tan\left(\frac{\alpha\pi}{2}\right)\left(\frac{\Delta\tau {Q^2}}{2}\right)^{\!\alpha}\frac{a_1^\alpha+a_2^\alpha}{2} \right]
  \right.
\\ 
     &
  \left.
     \quad \cdot           \exp \left[-\left(\frac{\Delta\tau {Q^2}}{2}\right)^{\!\alpha}\frac{a_1^\alpha+a_2^\alpha}{2} \right]
       \right\} \left(1+\epsilon Q\right)
 \; .
\end{split}
\end{equation}
 
Note that the cosine factor generates oscillations corresponding to alternating correlated and anti-correlated regions,
a feature clearly seen in the data (\Figs{fig:gauss_2jet} and~\ref{fig:gauss_3jet}).
Note also that since $a=1/\mt$ for two-jet events, the \taumodel\ predicts a decreasing effective source size with
increasing \mt.  These features are subjected to a quantitative test in the following sections.

\subsubsection{The \boldmath{$\tau$}-model for average \boldmath{$a$}}    \label{sect:tausimpl}
\begin{sidewaysfigure}
  \centering
  \includegraphics[width=.45\textwidth,bb=58 87 519 682,clip]{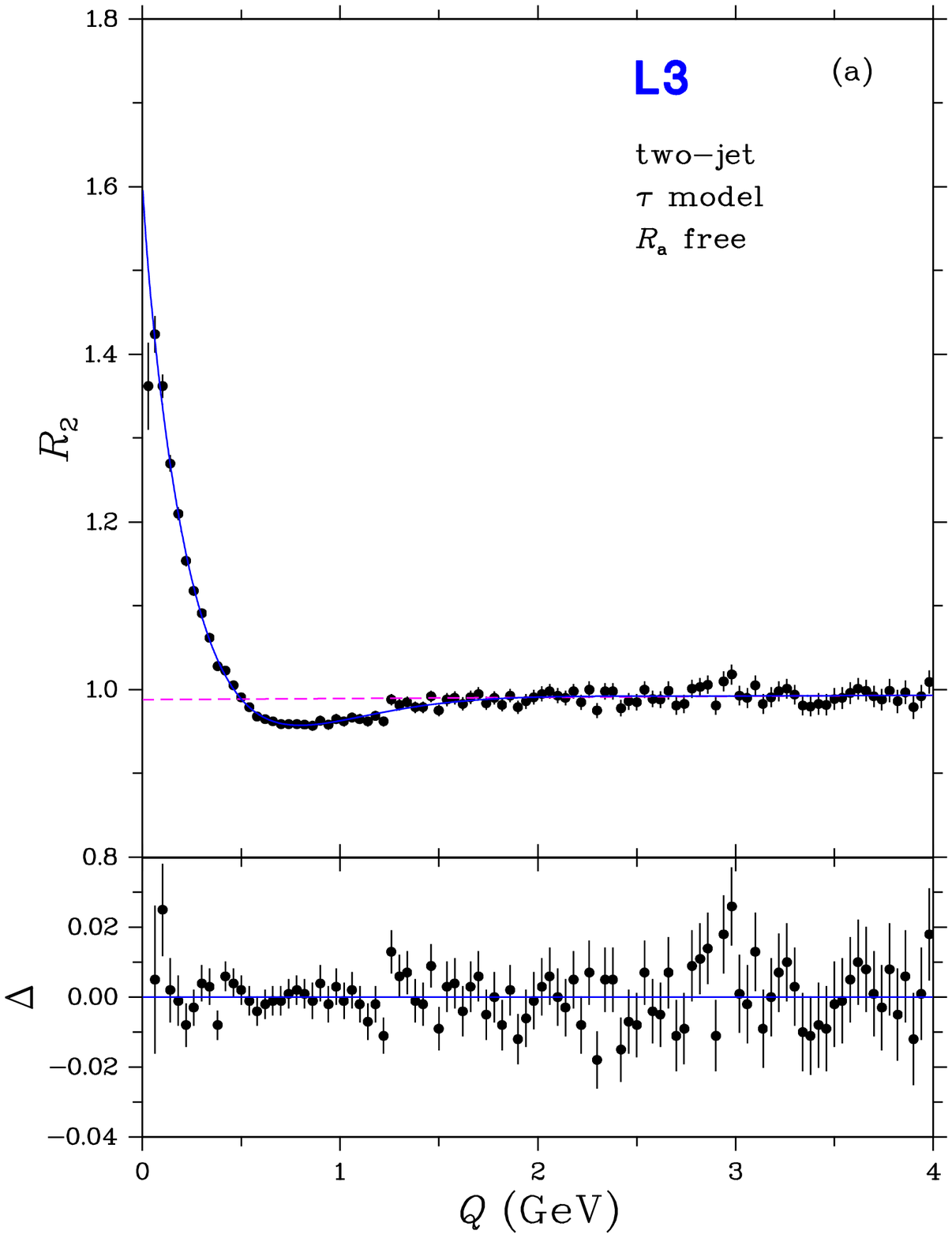}
  \includegraphics[width=.45\textwidth,bb=58 87 519 682,clip]{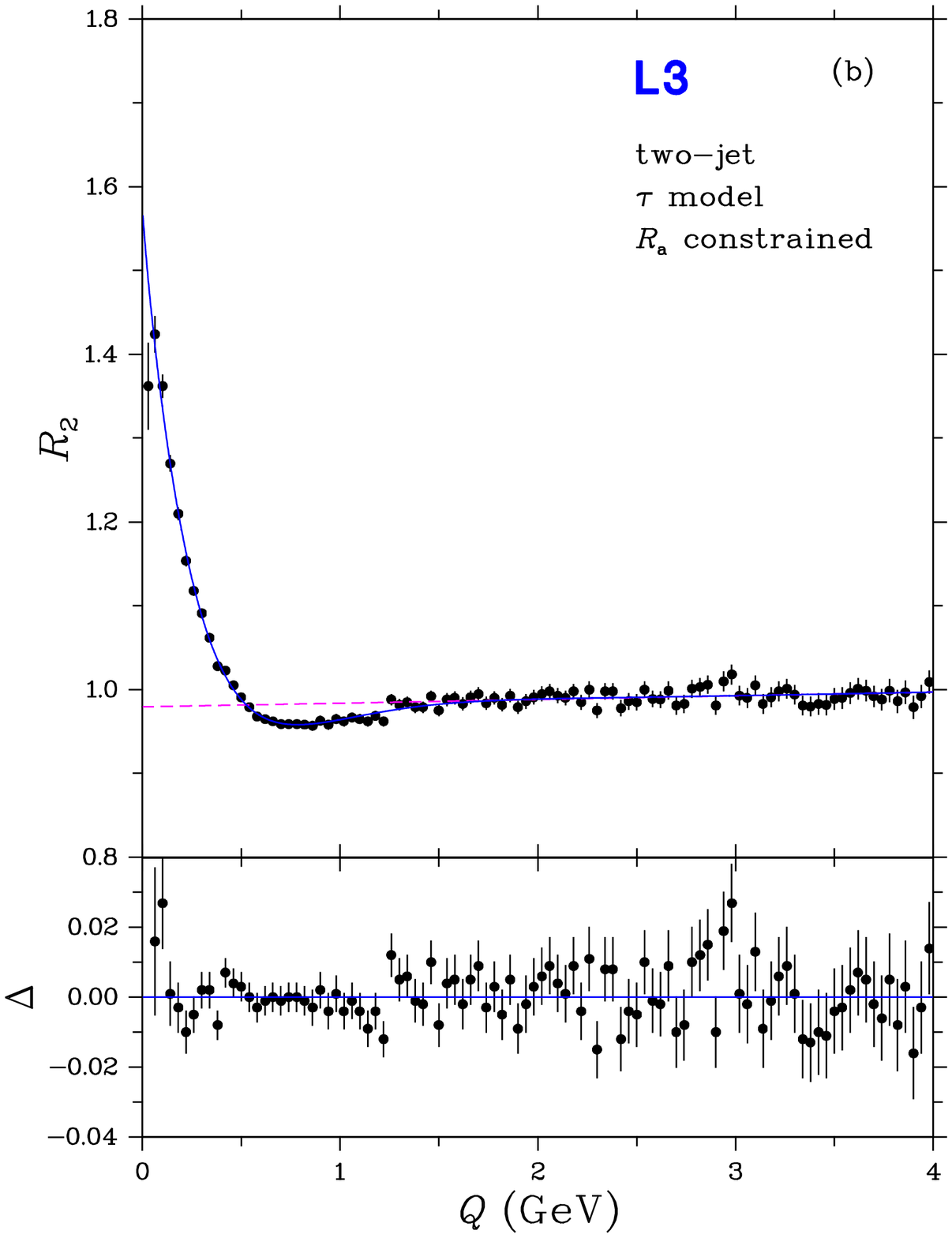}
  \caption{The Bose-Einstein correlation function $R_2$ for two-jet events. The curve
           corresponds to the fit of the one-sided L\'evy parametrization, \Eq{eq:asymlevR2},
           with the parameter $\Ra$ (a) free and (b) constrained by \Eq{eq:asymlevRaR}.
           The results of the fits are given in \Tabs{tab:a_levy} and \ref{tab:a_levy_c}, respectively.
           Also plotted is $\Delta$, the difference between the fit and the data.
           The dashed line represents the long-range part of the fit, \ie, $\gamma(1+\epsilon Q)$.
           \label{fig:a_levy_2jet}
           }
\end{sidewaysfigure}

\begin{sidewaysfigure}
  \centering
  \includegraphics[width=.45\textwidth,bb=58 87 519 682,clip]{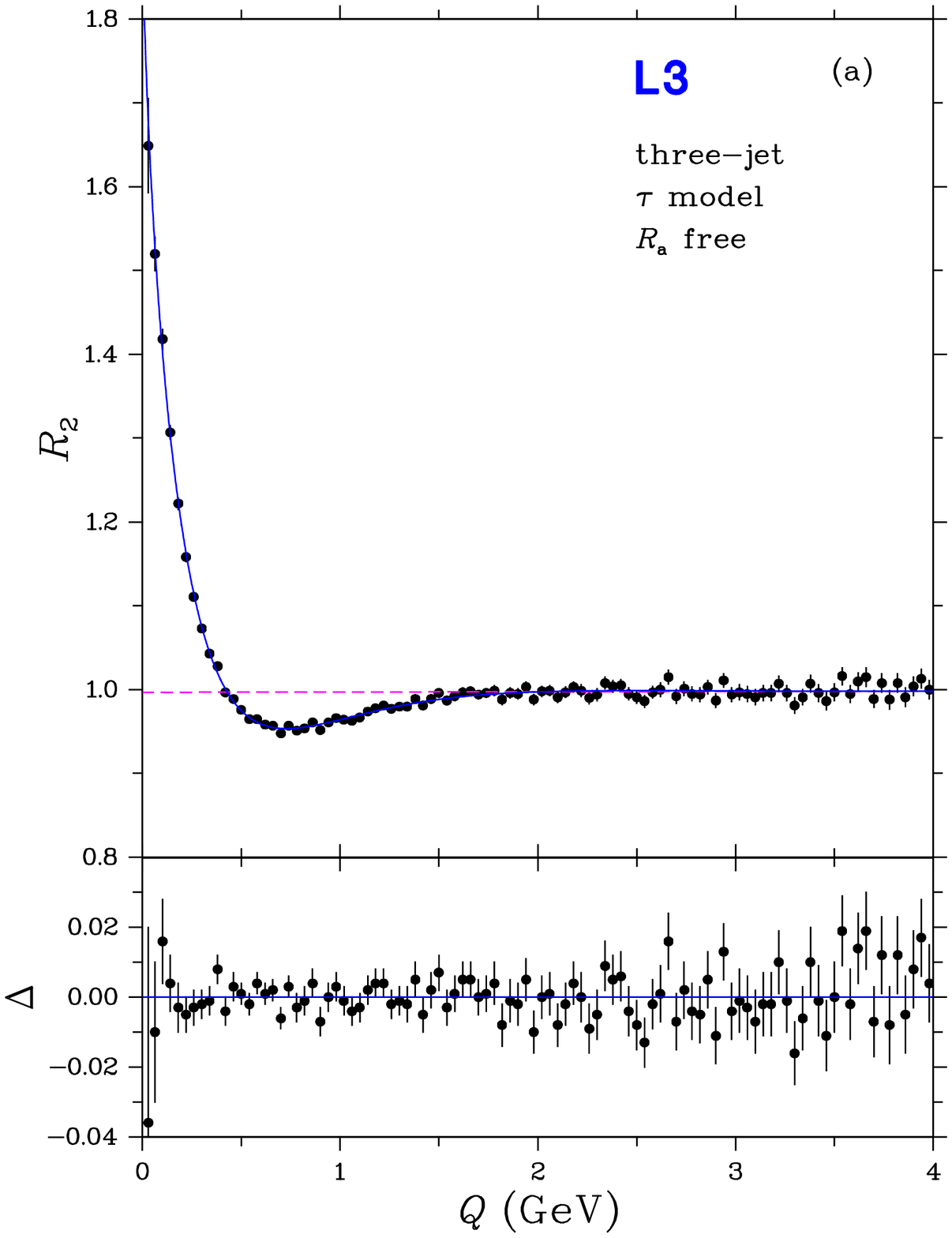}
  \includegraphics[width=.45\textwidth,bb=58 87 519 682,clip]{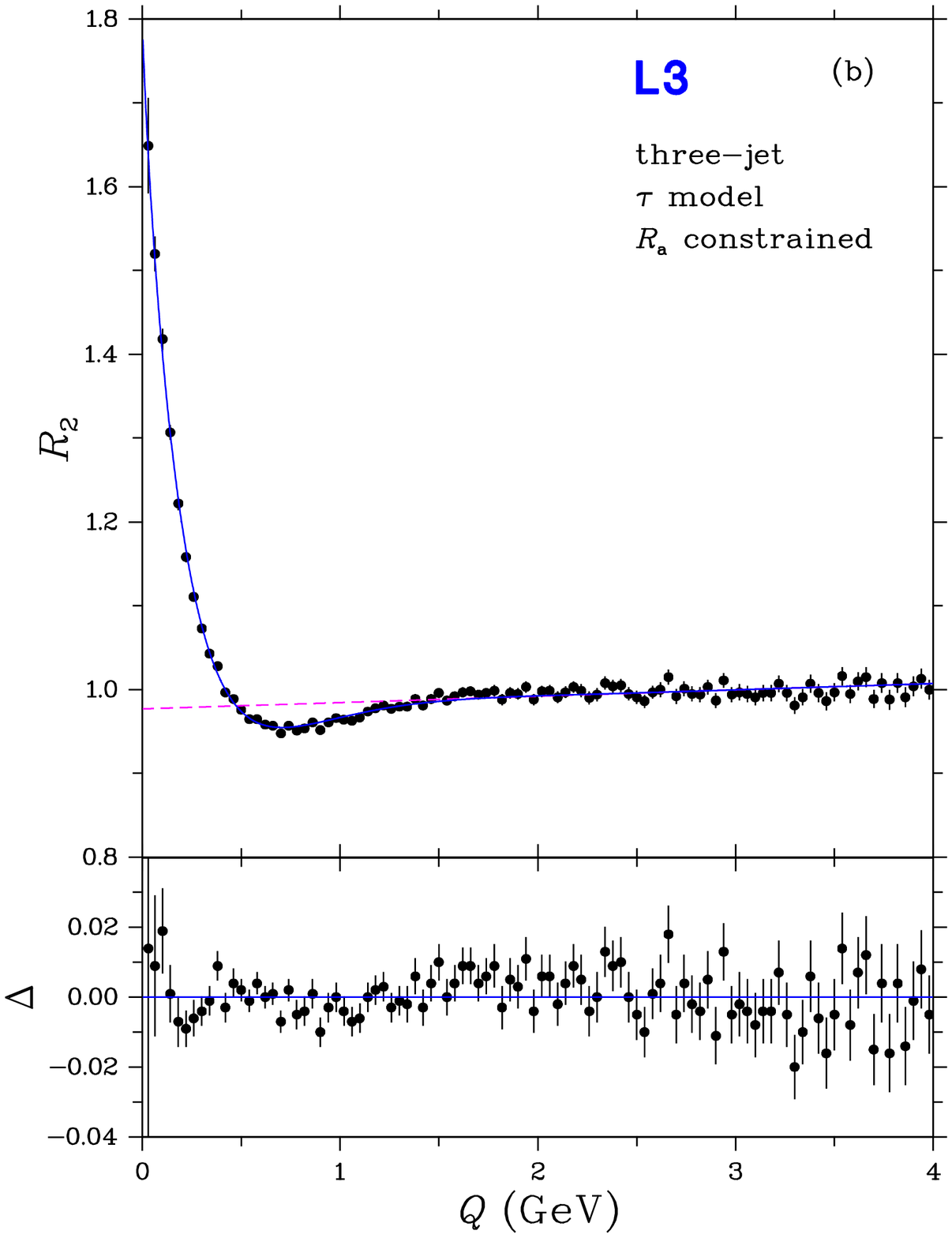}
  \caption{The Bose-Einstein correlation function $R_2$ for three-jet events. The curve
           corresponds to the fit of the one-sided L\'evy parametrization, \Eq{eq:asymlevR2},
           with the parameter $\Ra$ (a) free and (b) constrained by \Eq{eq:asymlevRaR}.
           The results of the fits are given in \Tabs{tab:a_levy} and \ref{tab:a_levy_c}, respectively.
           Also plotted is $\Delta$, the difference between the fit and the data.
           The dashed line represents the long-range part of the fit, \ie, $\gamma(1+\epsilon Q)$.
           \label{fig:a_levy_3jet}
           }
\end{sidewaysfigure}

Fits of \Eq{eq:levyR2a} are difficult since $R_2$ depends on three variables: $Q$, $a_1$ and $a_2$. 
Further, we have a simple expression for $a$ only for two-jet events.
Therefore, we first consider a simplification of \Eq{eq:levyR2a}
obtained by assuming (a) that particle production starts immediately, \ie, $\tau_0=0$,
and (b) an average $a$-dependence, which is implemented  
by introducing an effective radius defined by
\begin{equation}\label{eq:effR}
    R^{2\alpha} = \left(\frac{\Delta\tau}{2}\right)^{\!\alpha} \frac{a_1^\alpha+a_2^\alpha}{2} \;.
\end{equation}
This results in
\begin{equation}\label{eq:asymlevR2}
    R_2(Q) = \gamma \left[ 1+ \lambda \cos \left(\left(R_\mathrm{a}Q\right)^{2\alpha} \right)
             \exp \left(-\left(RQ\right)^{2\alpha} \right) \right] (1+ \epsilon Q) \;,
\end{equation}
where $R_\mathrm{a}$ is  related to $R$ by
\begin{equation}\label{eq:asymlevRaR}
    R_\mathrm{a}^{2\alpha} = \tan\left(\frac{\alpha\pi}{2}\right) R^{2\alpha} \;.
\end{equation}
Fits of
\Eq{eq:asymlevR2} are first performed with $R_\mathrm{a}$ as a free parameter.
The fit results obtained for two- and three-jet events are listed in \Tab{tab:a_levy}
and shown  in \Fig{fig:a_levy_2jet} for two-jet events and in \Fig{fig:a_levy_3jet} for three-jet events.
They have acceptable confidence levels, describing well the dip below unity in the 0.6--1.5\,\GeV\ region,
as well as the peak at low values of $Q$.

As shown in \Tab{tab:a_levy}, the estimates of some fit parameters are rather highly correlated.
Taking these correlations into account, the fit parameters for the two-jet events satisfy \Eq{eq:asymlevRaR},
the difference between the left- and right-hand sides of the equation being less than a standard deviation.
For three-jet events the difference amounts to about 1.5 standard deviations.
 
Note that no significant long-range correlation is observed: $\epsilon$ is zero within 1 standard deviation,
and fits with $\epsilon$ fixed to zero find the same values, within 1 standard deviation, of the parameters
as the fits of \Tab{tab:a_levy}.
Thus the method to remove non-Bose-Einstein correlations in the reference sample is apparently successful.
 
Fit results imposing \Eq{eq:asymlevRaR} are given in \Tab{tab:a_levy_c}.
For two-jet events, the values of the parameters are comparable to those with $\Ra$ free.
For three-jet events, the imposition of \Eq{eq:asymlevRaR} results in values of $\alpha$ and $R$
closer to those for two-jet events. 
Unlike the fits with \Ra\ free, the values of $\epsilon$ differ somewhat from zero, which could indicate a
slight deficiency in the description of BEC in the same way, though on a much smaller scale, as seen in the
Edgeworth and symmetric L\'evy fits of \Figs{fig:gauss_2jet} and \ref{fig:gauss_3jet}.
 
\begin{table}
\caption{Results of fits of \Eq{eq:asymlevR2}
           for two-jet and three-jet events,
           which are shown in \Figs{fig:a_levy_2jet}a and \ref{fig:a_levy_3jet}a, respectively.
         The first uncertainty is statistical, the second systematic (see Section \ref{sect:syst}).
\label{tab:a_levy}
         }
\begin{center}
\renewcommand{\arraystretch}{1.2}
\begin{tabular}{ l r@{$\;\pm\;$}l
                   r@{$\;\pm\;$}l
               }
\hline
    parameter               & \multicolumn{2}{c }{two-jet}        & \multicolumn{2}{c }{three-jet}         \\
\hline
 $\lambda$                  & 0.63  & 0.03$^{ +0.08}_{ -0.35}$  &  0.92  & 0.05$^{+0.06}_{-0.48}$    \\
 \rule{0pt}{11pt}$\alpha$   & 0.41  & $0.02^{ +0.04}_{ -0.06}$  &  0.35  & 0.01$^{+0.03}_{-0.04}$    \\
 $R$            (fm)        & 0.79  & $0.04^{+0.09}_{-0.19}$    &  1.06  & 0.05$^{+0.59}_{-0.31}$    \\
 $R_\mathrm{a}$ (fm)        & 0.69  & 0.04$^{+0.21}_{-0.09}$    &  0.85  & 0.04$^{+0.15}_{-0.05}$    \\
 $\epsilon$ (\invGeV)         & 0.001 & 0.002$^{+0.005}_{-0.008}$ &  0.000 & 0.002$^{+0.001}_{-0.007}$ \\
    $\gamma$                & 0.988 & 0.005$^{+0.026}_{-0.012}$ &  0.997 & 0.005$^{+0.019}_{-0.002}$  \\
\hline
  \chisq/DoF                & \multicolumn{2}{c }{91/94} & \multicolumn{2}{c }{84/94} \\
  confidence level          & \multicolumn{2}{c }{57\%}  & \multicolumn{2}{c }{76\%}  \\
\end{tabular}
\begin{tabular}{lccccc}
\hline
  \multicolumn{4}{l}{correlation coefficients}&\multicolumn{2}{c}{two-jet} \\
\hline
             & $\alpha$ & $R$        &   $R_\mathrm{a}$ &  $\epsilon$  & $\gamma$ \\
$\lambda$    & $-0.72$  & $\phm0.95$ & $\phm0.52$       &  $   -0.22$ & $\phm0.27$  \\
$\alpha$     &          & $   -0.62$ & $   -0.92$       &  $\phm0.72$ & $   -0.79$  \\
$R$          &          &            & $\phm0.38$       &  $   -0.10$ & $\phm0.13$  \\
$R_\mathrm{a}$&         &            &                  &  $   -0.89$ & $\phm0.94$  \\
$\epsilon$   &          &            &                  &              & $   -0.97$  \\
\hline
\end{tabular}
\
\begin{tabular}{lccccc}
\hline
  \multicolumn{4}{l}{\                       }&\multicolumn{2}{c}{three-jet} \\
\hline
             & $\alpha$ & $R$        &   $R_\mathrm{a}$ &  $\epsilon$  & $\gamma$ \\
$\lambda$    & $-0.78$  & $\phm0.97$ & $\phm0.65$       &  $   -0.32$ & $\phm0.38$  \\
$\alpha$     &          & $   -0.77$ & $   -0.95$       &  $\phm0.74$ & $   -0.81$  \\
$R$          &          &            & $\phm0.61$       &  $   -0.27$ & $\phm0.32$  \\
$R_\mathrm{a}$&         &            &                  &  $   -0.88$ & $\phm0.93$  \\
$\epsilon$   &          &            &                  &             & $   -0.98$  \\
\hline
\end{tabular}
\end{center}
\end{table}
 
\begin{table}
\caption{Results of fits of \Eq{eq:asymlevR2} imposing \Eq{eq:asymlevRaR}
           for two-jet and three-jet events,
           which are shown in \Figs{fig:a_levy_2jet}b and \ref{fig:a_levy_3jet}b, respectively.
         The first uncertainty is statistical, the second systematic (see Section \ref{sect:syst}).
\label{tab:a_levy_c}
         }
 
\begin{center}
\renewcommand{\arraystretch}{1.2}
\begin{tabular}{ l r@{$\;\pm\;$}l
                   r@{$\;\pm\;$}l
               }
\hline
    parameter              & \multicolumn{2}{c }{two-jet}        & \multicolumn{2}{c }{three-jet} \\
\hline
 $\lambda$                 & 0.61  & $0.03^{+0.08}_{-0.26}$    &  0.84  & $0.04^{+0.04}_{-0.37}$    \\
 \rule{0pt}{11pt}$\alpha$  & 0.44  & $0.01^{+0.05}_{-0.02}$    &  0.42  & $0.01^{+0.02}_{-0.04}$   \\
 $R$            (fm)       & 0.78  & $0.04^{+0.09}_{-0.16}$    &  0.98  & $0.04^{+0.55}_{-0.14}$   \\
 $\epsilon$ (\invGeV)        & 0.005 & $0.001\pm0.003$           &  0.008 & $0.001\pm0.005$          \\
    $\gamma$               & 0.979 & $0.002^{+0.009}_{-0.003}$ &  0.977 & $0.001^{+0.013}_{-0.008}$ \\
\hline
  \chisq/DoF               & \multicolumn{2}{c }{95/95} & \multicolumn{2}{c }{113/95} \\
  confidence level         & \multicolumn{2}{c }{49\%}  & \multicolumn{2}{c }{10\%}   \\
\end{tabular}
\begin{tabular}{lcccc}
\hline
  \multicolumn{3}{l}{correlation coefficients}&\multicolumn{2}{c}{two-jet} \\
\hline
             & $\alpha$ & $R$        &  $\epsilon$ & $\gamma$    \\
$\lambda$    & $-0.92$  & $\phm0.96$ &  $\phm0.29$ & $   -0.34$  \\
$\alpha$     &          & $   -0.96$ &  $   -0.35$ & $\phm0.42$  \\
$R$          &          &            &  $\phm0.21$ & $   -0.26$  \\
$\epsilon$   &          &            &             & $   -0.89$  \\
\hline
\end{tabular}
\
\begin{tabular}{lcccc}
\hline
  \multicolumn{3}{l}{\                       }&\multicolumn{2}{c}{three-jet} \\
\hline
             & $\alpha$ & $R$        &  $\epsilon$ & $\gamma$   \\
$\lambda$    & $-0.93$  & $\phm0.96$ &  $\phm0.24$ & $   -0.27$  \\
$\alpha$     &          & $   -0.97$ &  $   -0.31$ & $\phm0.37$  \\
$R$          &          &            &  $\phm0.20$ & $   -0.25$  \\
$\epsilon$   &          &            &             & $   -0.89$  \\
\hline
\end{tabular}
\end{center}
\end{table}

\subsubsection{The \boldmath{$\tau$}-model with \boldmath{$\mt$} dependence}  \label{sect:taufits}
 
\begin{figure}
  \centering
  \includegraphics[width=.94\textwidth,bb=58 87 519 682,clip]{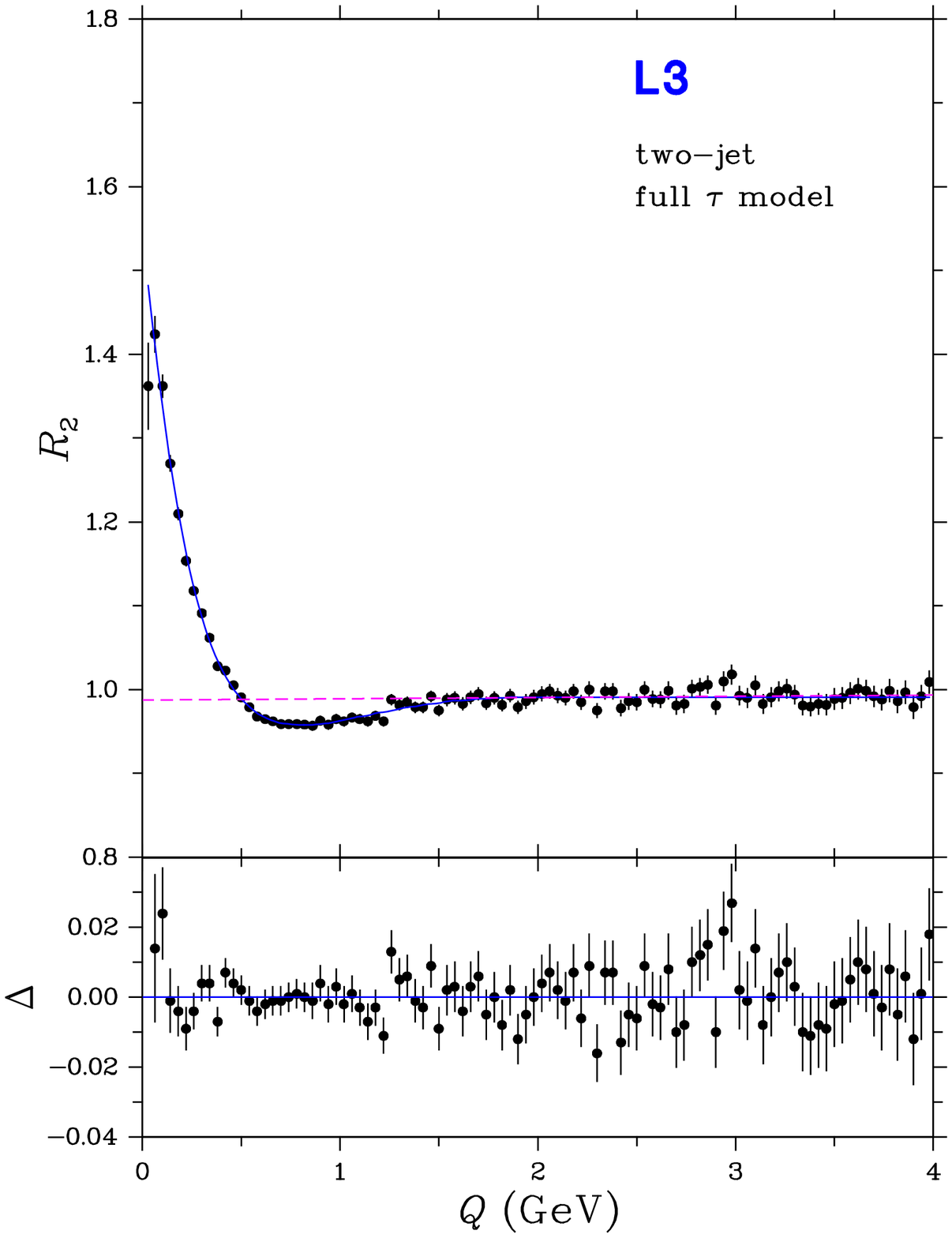}
  \caption{The Bose-Einstein correlation function $R_2$ for two-jet events. The curve
           corresponds to the fit of \Eq{eq:levyR2a}. 
           The results of this fit are given in \Tab{tab:2jetR2}.
           Also plotted is $\Delta$, the difference between the fit and the data.
           The dashed line represents the long-range part of the fit, \ie, $\gamma(1+\epsilon Q)$.
           \label{fig:2jetR2}
           }
\end{figure}

For two-jet events, $a=1/\mt$, while for three-jet events the situation is more complicated.
We therefore limit fits of  \Eq{eq:levyR2a} to the two-jet data.
For each bin in $Q$ the average values of $m_{\mathrm{t}1}$ and $m_{\mathrm{t}2}$ are calculated,
where $m_{\mathrm{t}1}$ and $m_{\mathrm{t}2}$ are the transverse masses of the
two particles making up a pair, requiring  $m_{\mathrm{t}1} > m_{\mathrm{t}2}$.
Using these averages,
\Eq{eq:levyR2a} is fit to $R_2(Q)$.
The fit results in $\tau_0=0.00\pm0.02$~fm, and we re-fit with $\tau_0$ fixed to zero.
The results are shown in \Fig{fig:2jetR2} and \Tab{tab:2jetR2}.
The parameters $\alpha$, $\Delta\tau$ and $\lambda$ are highly correlated.
\Eq{eq:levyR2a} for two-jet events can be simplified by requiring $m_{\mathrm{t}1}\approx m_{\mathrm{t}2}$.
This leads to fit results\cite{tamas:thesis} consistent with those without this simplification.

\begin{table}
\caption{Results of the fit of \Eq{eq:levyR2a} for two-jet events,
         which  is shown in \Fig{fig:2jetR2}.
         The parameter $\tau_0$ is fixed to zero.
         The first uncertainty is statistical, the second systematic (see Section \ref{sect:syst}).
\label{tab:2jetR2}
         }
\begin{center}
\renewcommand{\arraystretch}{1.2}
\begin{tabular}{ l r@{$\;\pm\;$}l
               }
\hline
    parameter               & \multicolumn{2}{c }{\ }           \\
\hline
  $\lambda$                 & 0.58  & $0.03^{+0.08}_{-0.24}$    \\
  \rule{0pt}{11pt}$\alpha$  & 0.47  & $0.01^{+0.04}_{-0.02}$    \\
  $\Delta\tau$   (fm)       & 1.56  & $0.12^{+0.32}_{-0.45}$    \\
  $\epsilon$ (\invGeV)        & 0.001 & $0.001\pm0.003$           \\
  $\gamma$                  & 0.988 & $0.002^{+0.006}_{-0.002}$ \\
\hline
  \chisq/DoF                & \multicolumn{2}{c }{90/95}        \\
  confidence level          & \multicolumn{2}{c }{62\%}         \\
\hline
\end{tabular}
 
\begin{tabular}{lcccc}
  \multicolumn{5}{l}{correlation coefficients} \\
\hline
             & $\alpha$ & $\Delta\tau$ &  $\epsilon$ & $\gamma$    \\
$\lambda$    & $-0.92$  & $\phm0.95$ &  $\phm0.30$ & $   -0.35$  \\
$\alpha$     &          & $   -0.96$ &  $   -0.37$ & $\phm0.45$  \\
$\Delta\tau$ &          &            &  $\phm0.23$ & $   -0.29$  \\
$\epsilon$   &          &            &             & $   -0.89$  \\
\hline
\end{tabular}
\end{center}
\end{table}

Since the \taumodel\ describes the \mt\ dependence of $R_2$, an important test of the model
is that its parameters, $\alpha$,  $\Delta\tau$, and $\tau_0$, not depend on \mt.
Note that the parameter $\lambda$, which is not a parameter of the \taumodel,
can depend on \mt, \eg, as a result of resonances~\cite{Bolz:1992hc,Csorgo:1994in}.
The large correlations between the fit estimates of $\lambda$ and those of the \taumodel\ parameters
complicate the testing of \mt-independence.  We perform fits in various regions of the $m_{\mathrm{t}1}$-$m_{\mathrm{t}2}$ plane
keeping $\alpha$ and $\Delta\tau$ fixed at the values obtained in the fit to the entire \mt\ plane (\Tab{tab:2jetR2}).
The regions are chosen such that the numbers of pairs of particles in the regions are comparable.
The \mt\ regions and the confidence levels of the fits are shown in \Tab{tab:2jetmtCL} along with the values of $\lambda$.
The confidence levels are seen to be quite reasonably distributed
in agreement with the hypothesis of \mt-independence of the parameters of the \taumodel\ ($\alpha$,  $\Delta\tau$, and $\tau_0$).
 
\begin{table}
\caption{Confidence levels and the values of $\lambda$ found in fits of \Eq{eq:levyR2a} for two-jet events
         in various  regions of the $m_{\mathrm{t}1}$-$m_{\mathrm{t}2}$ plane
         with $\alpha$ and  $\Delta\tau$                               fixed to the result of the fit to the entire plane.
\label{tab:2jetmtCL}
         }
\begin{center}
\begin{tabular}{cc|c|c|c|c}
\hline
    \multicolumn{2}{c|}{\mt\ regions (\GeV)}& \multicolumn{2}{c|}{average}  & confidence &            \\
    $m_{\mathrm{t}1}$  &  $m_{\mathrm{t}2}$ & \multicolumn{2}{c|}{\mt\ (\GeV)}& level    & $\lambda$  \\
                       &                    &$Q<0.4$&all&    (\%)         &           \\
\hline
    0.14 -- 0.26       &   0.14 -- 0.22     & 0.19 & 0.19 &  10    &$0.39\pm0.02$ \\ 
    0.14 -- 0.34       &   0.22 -- 0.30     & 0.27 & 0.27 &  48    &$0.76\pm0.03$ \\ 
    0.14 -- 0.46       &   0.30 -- 0.42     & 0.37 & 0.37 &  74    &$0.83\pm0.03$ \\ 
    0.14 -- 0.66       &   0.42 -- 4.14     & 0.52 & 0.52 &  13    &$0.97\pm0.04$ \\ 
    0.26 -- 0.42       &   0.14 -- 0.22     & 0.25 & 0.26 &  22    &$0.53\pm0.02$ \\ 
    0.34 -- 0.46       &   0.22 -- 0.30     & 0.32 & 0.33 &  33    &$0.80\pm0.03$ \\ 
    0.46 -- 0.58       &   0.30 -- 0.42     & 0.43 & 0.44 &  34    &$0.91\pm0.04$ \\ 
    0.66 -- 0.86       &   0.42 -- 4.14     & 0.65 & 0.65 &  66    &$1.01\pm0.05$ \\ 
    0.42 -- 0.62       &   0.14 -- 0.22     & 0.34 & 0.34 &  17    &$0.41\pm0.03$ \\ 
    0.46 -- 0.70       &   0.22 -- 0.30     & 0.41 & 0.41 &  55    &$0.64\pm0.03$ \\ 
    0.58 -- 0.82       &   0.30 -- 0.42     & 0.52 & 0.52 &  59    &$0.70\pm0.04$ \\ 
    0.86 -- 1.22       &   0.42 -- 4.14     & 0.80 & 0.81 &  24    &$0.66\pm0.05$ \\ 
    0.70 -- 4.14       &   0.22 -- 0.30     & 0.59 & 0.65 &  \pho4 &$0.37\pm0.04$ \\ 
    0.82 -- 4.14       &   0.30 -- 0.42     & 0.71 & 0.76 &  11    &$0.56\pm0.05$ \\ 
\hline
\end{tabular}
\end{center}
\end{table}
 
\subsection{Systematic Uncertainties} \label{sect:syst}
The following sources of systematic uncertainty are investigated:
\begin{itemize}
  \item{Event and track selection:} The cuts used in selecting hadronic events and high precision tracks are
varied within reasonable limits:
  (a) \Evis\ within $\sqrt{s}\pm0.45\sqrt{s}$ to within $\sqrt{s}\pm0.55\sqrt{s}$;
  (b)   transverse energy imbalance from $0.5\Evis$ to $0.7\Evis$
  and longitudinal energy imbalance from $0.3\Evis$ to $0.5\Evis$;
  (c) minimum number of calorimeter clusters from  13 to 17;
  (d) maximum value of $\abs{\cos(\Theta)}$ from 0.707 to 0.777 and
      maximum value of $\abs{\cos(\Theta_\mathrm{TEC})}$ from 0.662 to 0.736;
  (e) transverse momentum of a track from 100~\MeV\ to 200~\MeV;
  (f) minimum of 21 hits spanning at least 48 wires    to 31 hits spanning at least 32 wires;
  (g) maximum distance of closest approach from 5~mm to 15~mm.
 
  \item{Fit range:} The fits presented above were performed in the range $0<Q<4$\,\GeV.
  The fits were repeated in the ranges $0.04<Q<4$\,\GeV, $0<Q<3$\,\GeV\ and $0<Q<5$\,\GeV.
Note that uncertainty on the contribution of FSI is included by
considering the range $0.04<Q<4$\,\GeV, FSI being most important at the smallest values of $Q$.
 
  \item{Mixing:} The event mixing method to construct $\rho_0$ uses tracks from events having similar
multiplicity.   The definition of ``similar'' was varied by changing the range of multiplicity classes from
which the mixed tracks are chosen from 0 to 8, \ie, from demanding the same multiplicity class as the data
event to including classes within $\pm4$ of that of the data event.

  \item{Monte Carlo:} Events generated by \JETSET\ with no BEC simulation or by \HERWIG\ \cite{HERWIG59} are used
instead of \JETSET\ with BEC for the correction of $\rho_2$ for detector acceptance,  efficiency and non-pion pair background
and \PYTHIA\ \cite{PYTHIAsix} and \HERWIG\ \cite{HERWIGsix} are used\footnote{\HERWIG\ was modified to
use the decay and BEC routines of \PYTHIA. Fragmentation parameters of both \HERWIG\ and \PYTHIA\ were
tuned using \Lthreefoot\ data with BEC simulated by the \BEtt\ Gaussian algorithm \cite{PYBOEI}.
The events are generated using these parameters but without BEC simulation.}
instead of \JETSET\ without BEC to correct $\rho_0$.
 
\end{itemize}
For each source the root mean square of the deviations from the values measured in the standard selection is taken
as the systematic uncertainty, positive and negative deviations being treated separately.
The systematic uncertainties from the four sources are added in quadrature to obtain the total systematic uncertainty.
For the physically interesting parameters ($\alpha$, $R$, \Ra, $\Delta\tau$), the largest contributions come from Monte Carlo
variation and/or mixing. The other parameters sometimes receive significant contributions from fit range and/or track and event
selection.

\section{The emission function of two-jet events}  \label{sect:emission2jet}
Within the framework of the \taumodel, we now
reconstruct the space-time picture of the emitting process for two-jet events.
The emission function in space-time,
 $S_{\mathrm{r}\tau}(\vec{r},\tau)$,
normalized to unity,
is given in the \taumodel\ by  \cite{ourTauModel}:
\begin{equation}   \label{eq:Sspacetau}
   S_{\mathrm{r}\tau}(\vec{r},\tau) = \nbarinv\frac{\mathrm{d}^4 n}{\dd{\tau}\dd{\vec{r}}}
        = \nbarinv \left(\frac{\mt}{\tau}\right)^{\!3} H(\tau) \rho_\mathrm{p}\left(\vec{p}=\frac{\mt\vec{r}}{\tau} \right) \;,
\end{equation}
where $n$ and $\bar{n}$ are, respectively, the number and average number of pions produced, and
where $\rho_\mathrm{p}(\vec{p})$ is the single-particle momentum distribution,
\begin{equation}   \label{eq:rho1}
   \rho_\mathrm{p}(\vec{p}) =   \frac{\mathrm{d}^3 n}{\dd{p_\mathrm{x}}\dd{p_\mathrm{y}}\dd{\pz}} \;,
\end{equation}
which is normalized to the mean multiplicity, $\bar{n}$.
Note that, given the \taumodel\ correlation between space-time and momentum space,
\begin{equation}   \label{eq:mt}
  \mt = \frac{m\tau}{\sqrt{\tau^2-(\rx^2+\ry^2)}} \;.
\end{equation}
 Using $\tau^2=t^2-\rz^2$ and \Eq{eq:mt} , we can rewrite \Eq{eq:Sspacetau} in terms of $\vec{r}$
and $t$:
\begin{equation}   \label{eq:Sspacet}
   S_{\mathrm{rt}}(\vec{r},t) = \nbarinv\frac{\mathrm{d}^4 n}{\dd{t}\dd{\vec{r}}}
        = \nbarinv J_{\mathrm{rt}}(\vec{r},t) \; H\left(\tau=\sqrt{t^2-\rz^2}\right)
           \rho_\mathrm{p}\left(\vec{p}=\frac{m\vec{r}}{\sqrt{t^2-(\rx^2+\ry^2+\rz^2))}} \right) \;,
\end{equation}
where
\begin{equation}   \label{eq:SspacetJ}
   J_{\mathrm{rt}}(\vec{r},t) =
  \frac{m^3t}{\left(t^2-\left(\rx^2+\ry^2+\rz^2\right)\right)^{3/2}\left(t^2-\rz^2\right)^{1/2}}
\end{equation}
is the Jacobian of the variable transformation.

Given the symmetry of two-jet events, it is convenient to write $S$
in cylindrical coordinates and average over the azimuthal angle.
To simplify the reconstruction of $S$ we assume that
the momentum distribution $\rho_\mathrm{p}$ can be factorized in the product of transverse and longitudinal distributions, \eg,
  $   \rho_\mathrm{y\pt} = \rho_\mathrm{y}(y) \rho_\mathrm{\pt}(\pt)$,
where $y$ is the rapidity and \pt\ is the transverse momentum.
This is found to be a reasonable approximation except at very  high values of $\abs{y}$ or \pt\ \cite{tamas:thesis}.
 
\begin{figure}
  \centering
   \includegraphics*[width=.6\figwidth,angle=-90,viewport=100 0 567 765]{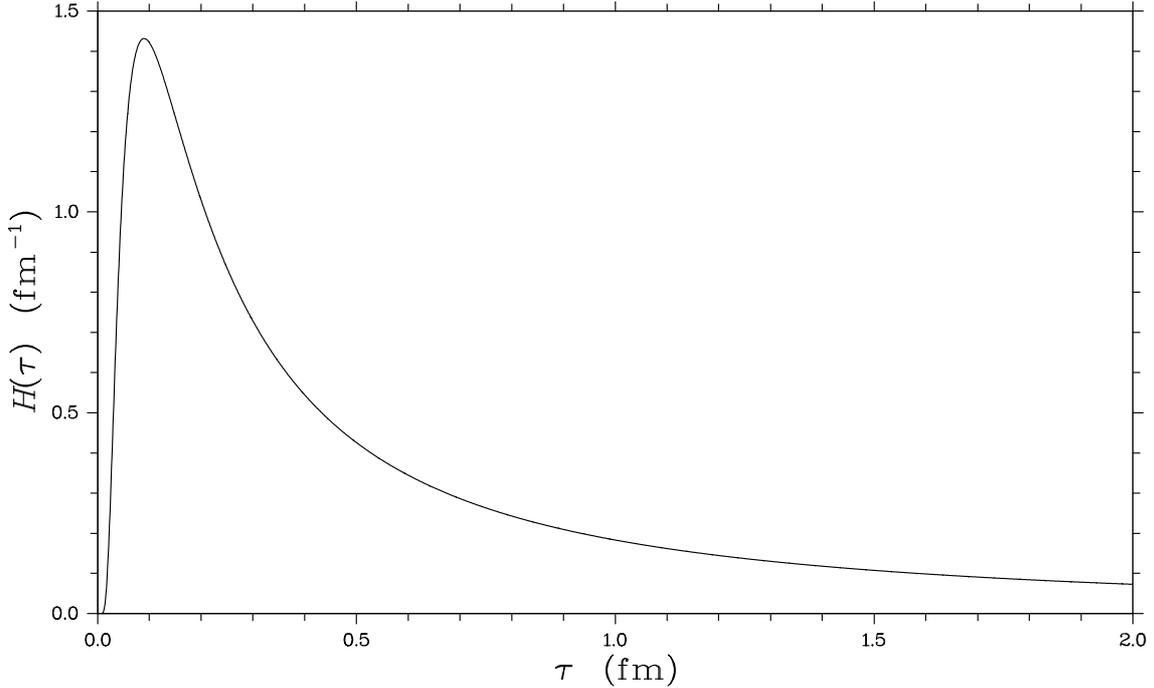}
  \caption{The proper-time distribution, $H(\tau)$,
           for $\alpha=0.47$, $\tau_0=0$ and $\Delta\tau=1.56$\,fm.
           $H(\tau)$ was calculated using the program \texttt{STABLE} \cite{nolan:stable}.
  \label{fig:Htau}
  }
\end{figure}

\begin{figure}
  \centering
   \includegraphics*[width=.49\figwidth,bb=24 9 515 370]{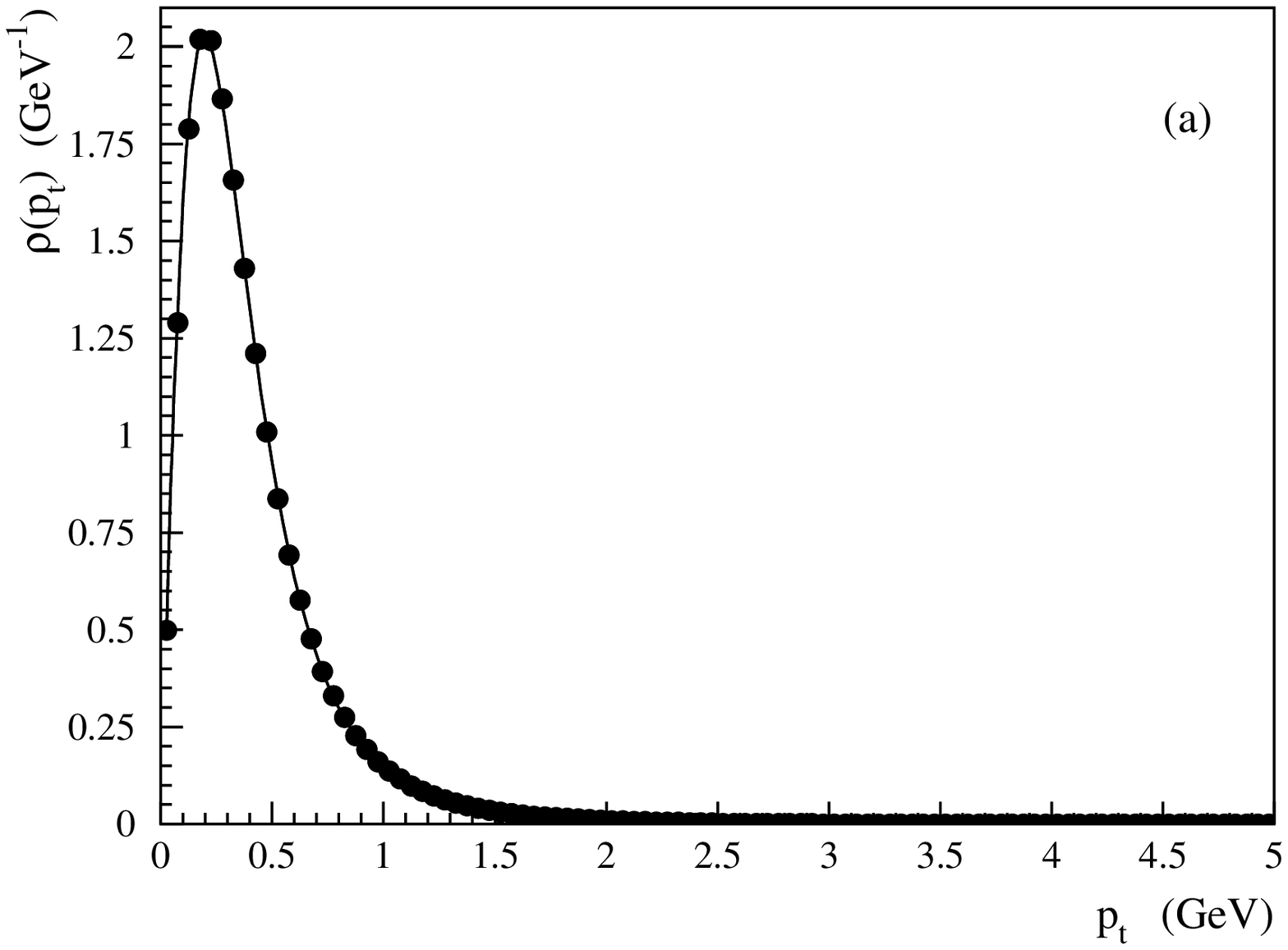}
   \hfill
   \includegraphics*[width=.49\figwidth,bb=24 9 515 370]{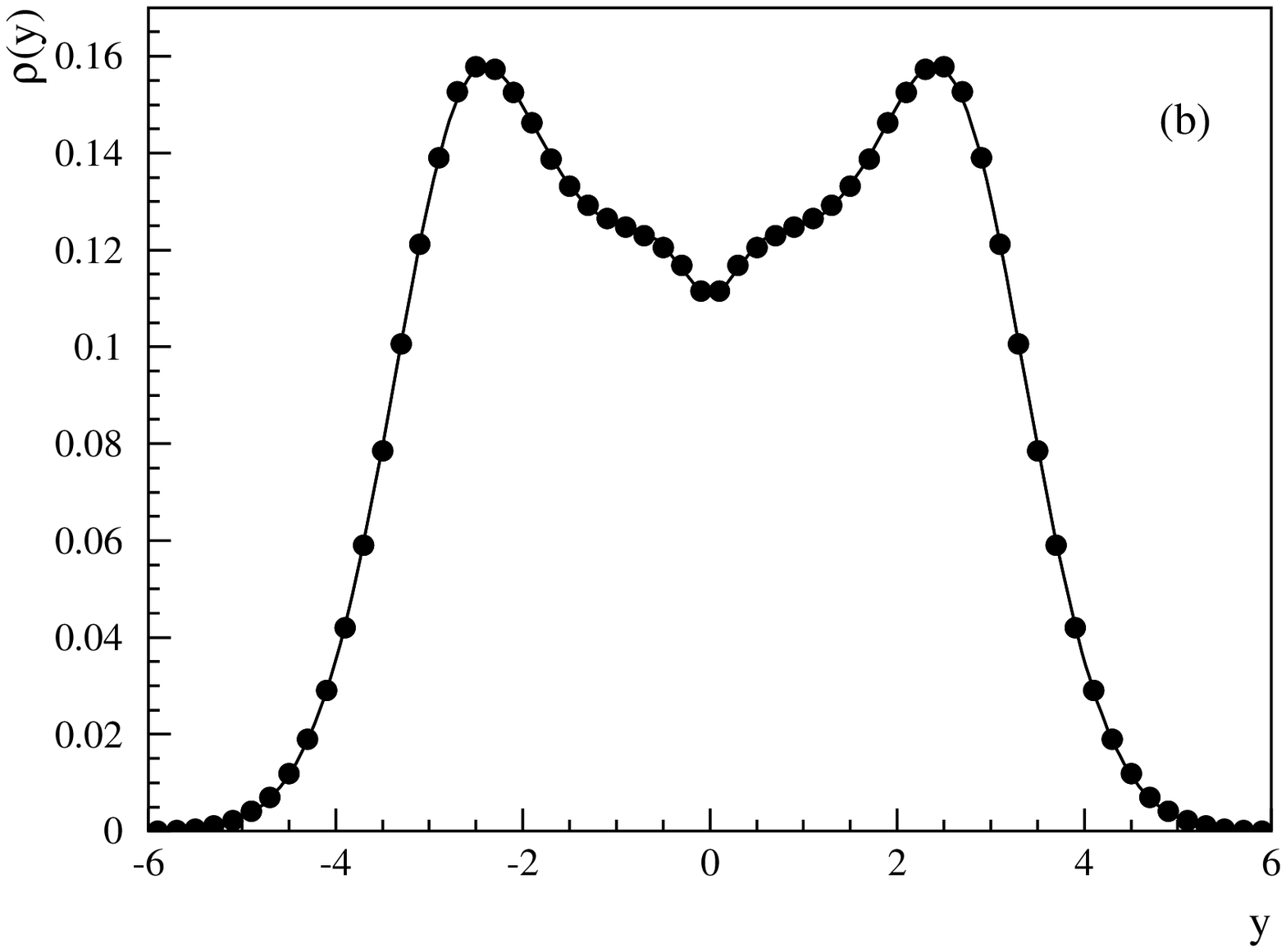}
  \caption{The (a) \pt\ and (b) rapidity distributions of two-jet events and the parametrizations used in the
           reconstruction of the source function.
  \label{fig:yptparam}
  }
\end{figure}

Using
$H(\tau)$ as obtained from the fit of \Eq{eq:levyR2a} 
(\Tab{tab:2jetR2}),
which is shown in \Fig{fig:Htau},
together with the inclusive \pt\ and rapidity distributions,%
\footnote{These distributions of high precision 
charged tracks from \Lthreefoot\ two-jet events are corrected for detector acceptance and efficiency bin-by-bin
by the ratio of the distributions of generator-level to detector-level Monte Carlo samples.
For the purpose of the reconstruction, the \pt\ and rapidity
distributions 
have been parametrized as 
$\rho_\mathrm{\pt}(\pt)=\exp\left(-8.340\pt + 1.249\pt^2 -0.1050\pt^3\right)
       \left(19.52\pt + 182.5\pt^2 -144.4\pt^3 + 90.37\pt^4 +48.84\pt^5\right)$
and
$\rho_\mathrm{y}(y)=      \exp\left(-\left[(-0.621-\abs{y}) / 1.605\right]^2\right) $
     $ \left(0.1272\right.  + 0.07338 \abs{y}  + 0.2155 y^2  - 0.3577 \abs{y^3}  + 0.4955 y^4 $
     $ - 0.2843\abs{y^5}     \left. + 0.07887 y^6 \right)$.}
which are shown in \Fig{fig:yptparam},
the full emission function is reconstructed.

\begin{figure}
  \centering
  \includegraphics*[width=.49\figwidth]{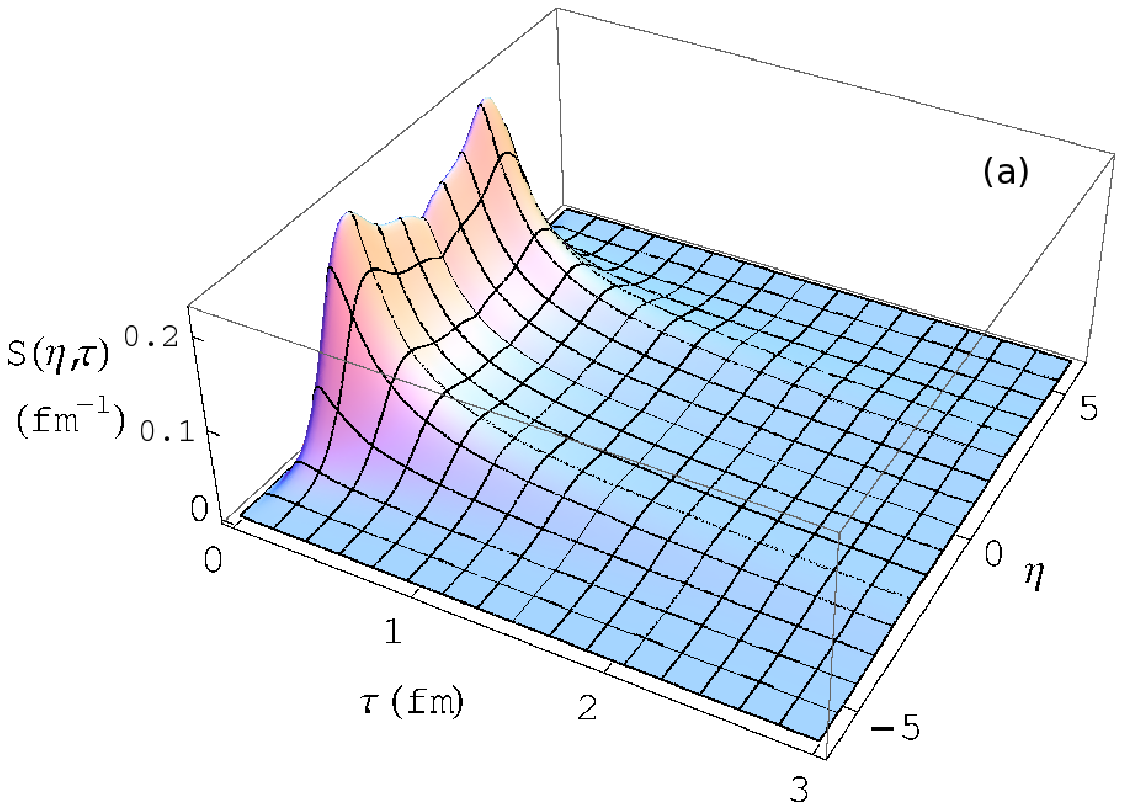}
        \hfil
  \includegraphics*[width=.49\figwidth]{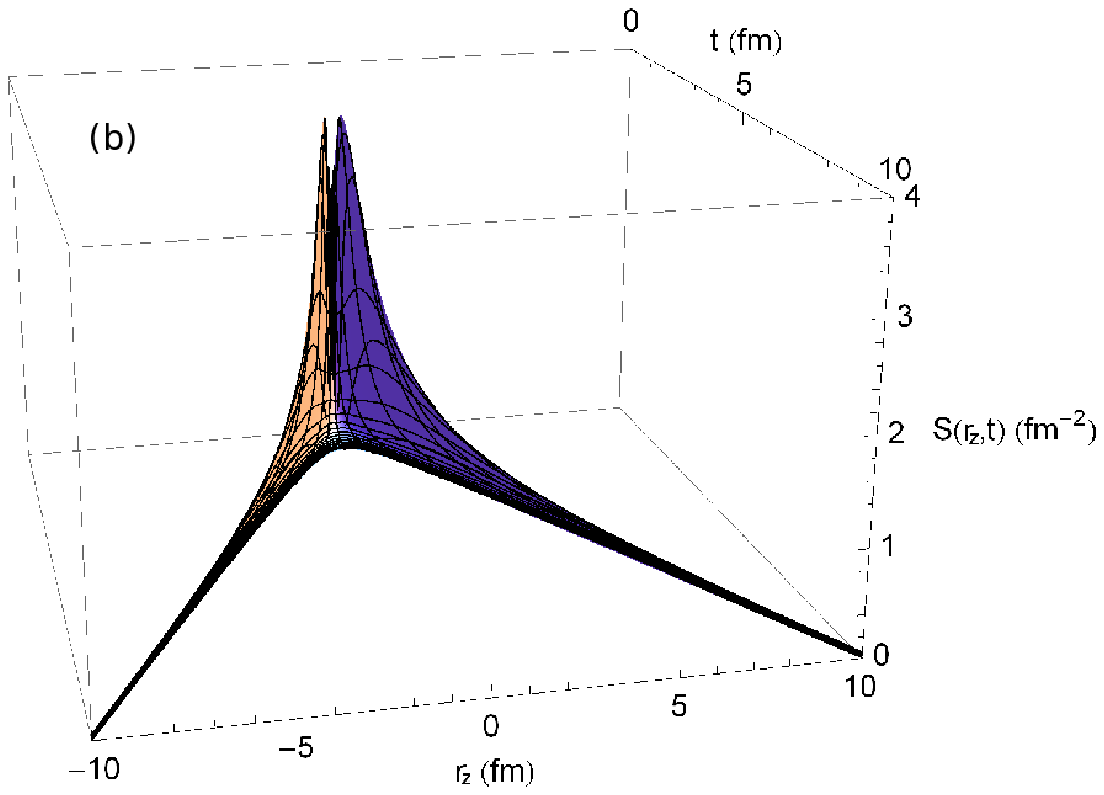}  \\ 
\hfil
  \caption{The temporal-longitudinal part of the source function of two-jet events,
           (a) $S(\eta,\tau)$, \Eq{eq:Setatau} and
           (b) $S(\rz,t)$, \Eq{eq:Szt},
           normalized to  unity.
\label{fig:longemis}
          }
\end{figure}

Integrating \Eqs{eq:Sspacetau}  and (\ref{eq:Sspacet}) over the transverse coordinates results, respectively,
in
\begin{equation}   \label{eq:Setatau}
   S_{\eta\tau}(\eta,\tau) = \nbarinv\frac{\mathrm{d}^2 n}{\dd{\eta}\dd{\tau}}
        = \nbarinv                         H\left(\tau\right)
           \rho_\mathrm{y}\left(y=\eta\right)
\end{equation}
and
\begin{equation}   \label{eq:Szt}
   S_{\mathrm{zt}}(\rz,t) = \nbarinv\frac{\mathrm{d}^2 n}{\dd{\rz}\dd{t}}
        = \nbarinv\frac{1}{\sqrt{t^2-\rz^2}} H\left(\tau=\sqrt{t^2-\rz^2}\right)
           \rho_\mathrm{y}\left(y=\eta=\frac{1}{2}\ln\frac{t+r_\mathrm{z}}{t-\rz}  \right) \;.
\end{equation}
These are plotted in \Fig{fig:longemis}.
They exhibit a ``boomerang'' shape with maxima at low values of $\tau$ and $\eta$ or $t$ and $\rz$,
but with tails reaching out to very large values,
a feature also observed in hadron-proton~\cite{NA22emiss} and heavy ion collisions~\cite{HIemiss:Ster}.
        Note however, that in the case of two-jet events, the emission function
        $S_{\eta \tau}(\eta,\tau)$ has two maxima, as does
        $S_{zt}(\rz,t)$, as is seen  in \Fig{fig:longemis}.
        These two different maxima correspond to the two-jet structure of
        the events. This is in marked contrast with the reconstructed
        emission  function in hadron-proton reactions
        at $\sqrt{s} = 22\;\GeV$, where only a single maximum is observed\cite{NA22emiss}.

The transverse part of the emission function is obtained by integrating \Eq{eq:Sspacetau}  over \rz\
and averaging over the azimuthal angle:
\begin{equation}   \label{eq:Sxytau}
   S_{\mathrm{xy}\tau}(\rx,\ry,\tau) = \nbarinv\frac{\mathrm{d}^3 n}{\dd{\rx}\dd{\ry}\dd{\tau}}
        = \nbarinv J_{\mathrm{xy}\tau}(\rx,\ry,\tau) \;
              H(\tau)
\rho_\mathrm{\pt}\left(\pt=\frac{m\sqrt{\rx^2+\ry^2\strut}}{\sqrt{\tau^2-(\rx^2+\ry^2)}} \right) \;,
\end{equation}
where
\begin{equation}   \label{eq:SxytauJ}
   J_{\mathrm{xy}\tau}(\rx,\ry,\tau) = \frac{m}{2\pi\sqrt{\rx^2+\ry^2\strut}\sqrt{\tau^2-(\rx^2+\ry^2})} \;.
\end{equation}
 
\begin{figure}
  \centering
   \includegraphics[width=.49\figwidth,bb=14 14 335 247,clip]{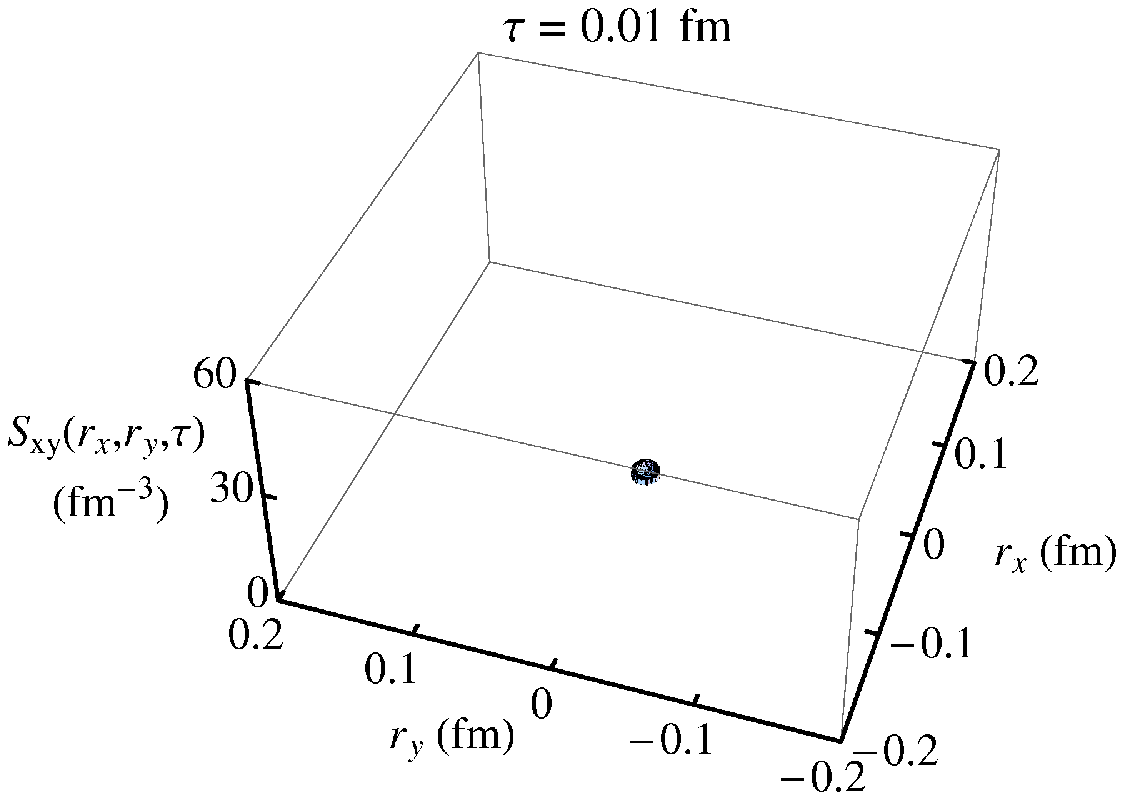}
   \includegraphics[width=.49\figwidth,bb=14 14 335 247,clip]{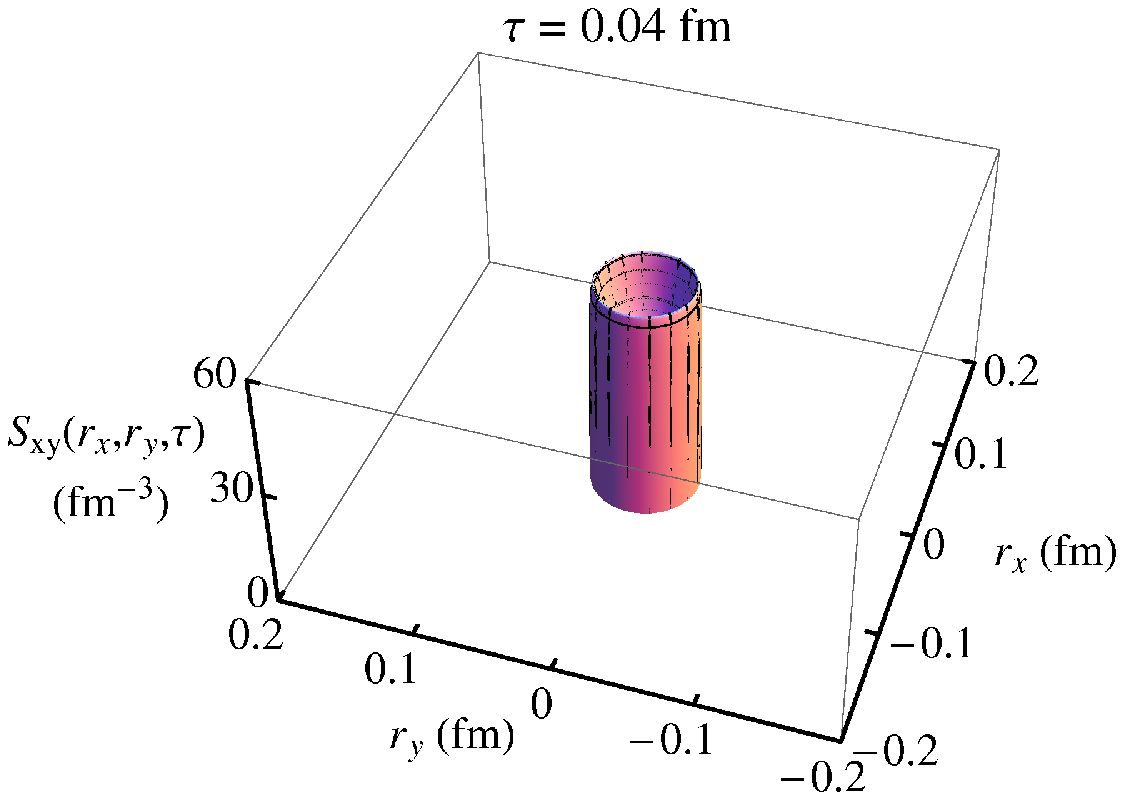}
   \includegraphics[width=.49\figwidth,bb=14 14 335 247,clip]{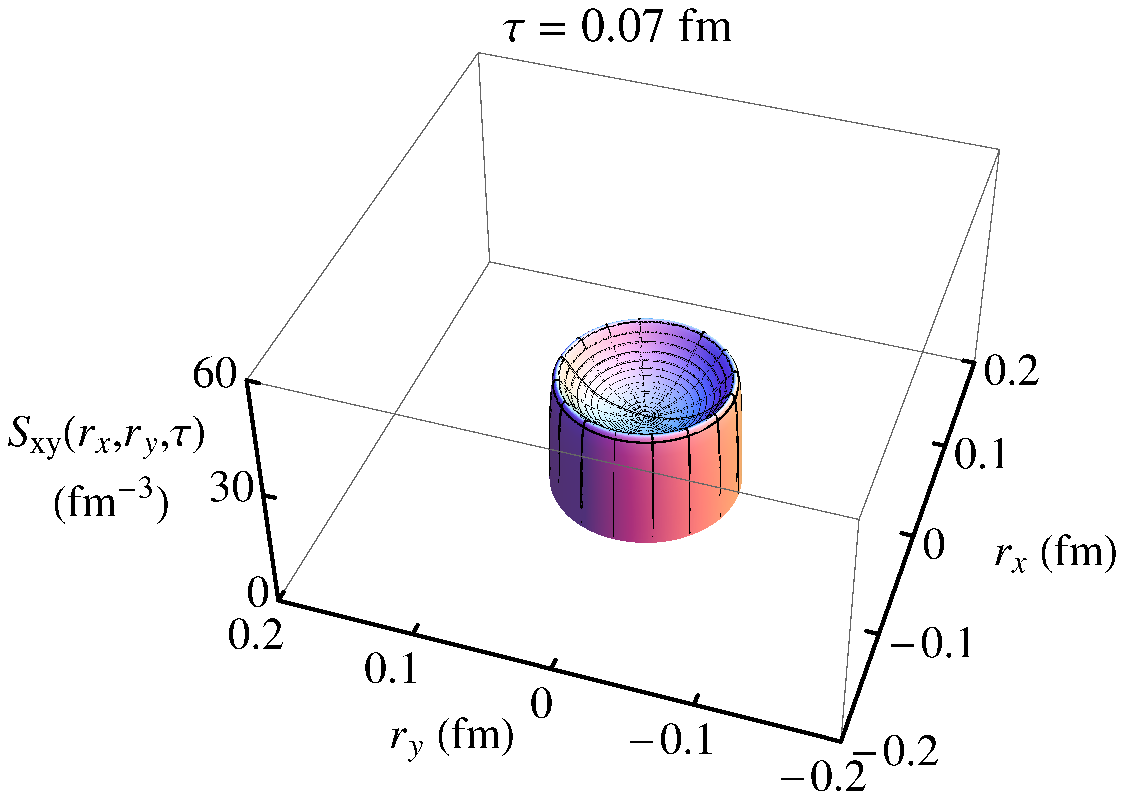}
   \includegraphics[width=.49\figwidth,bb=14 14 335 247,clip]{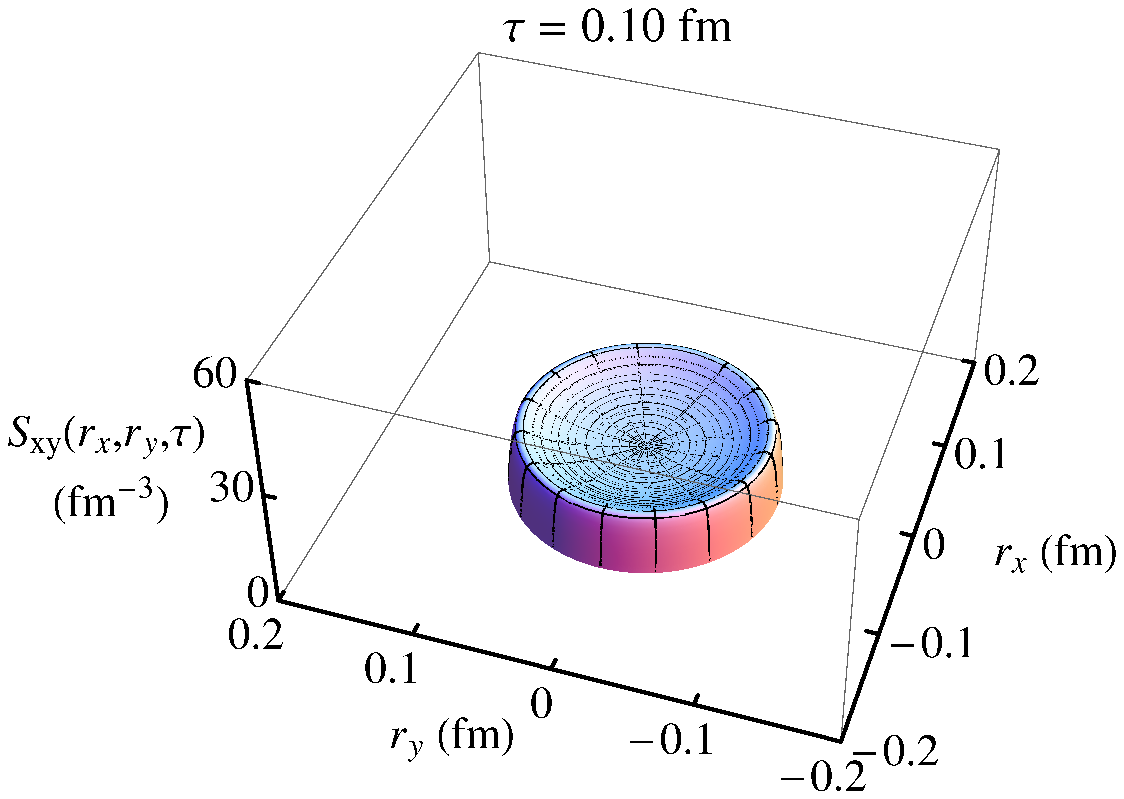}
   \includegraphics[width=.49\figwidth,bb=14 14 335 247,clip]{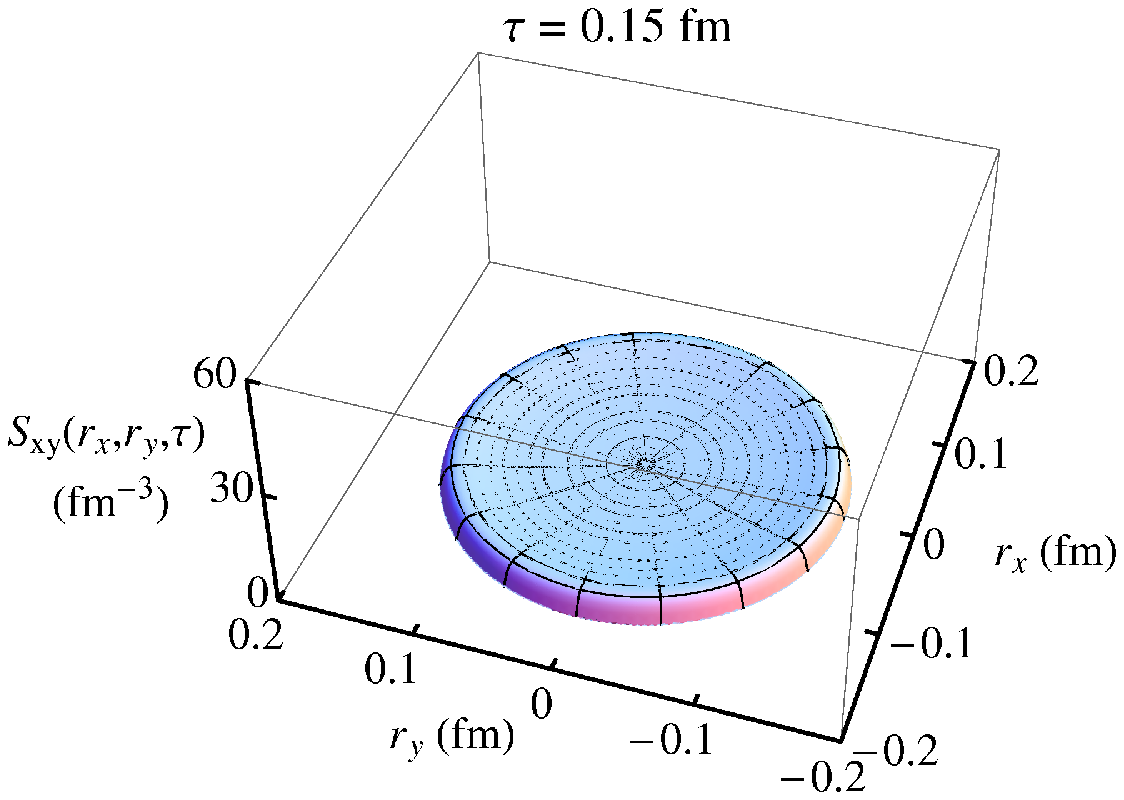}
   \includegraphics[width=.49\figwidth,bb=61 391 526 681,clip]{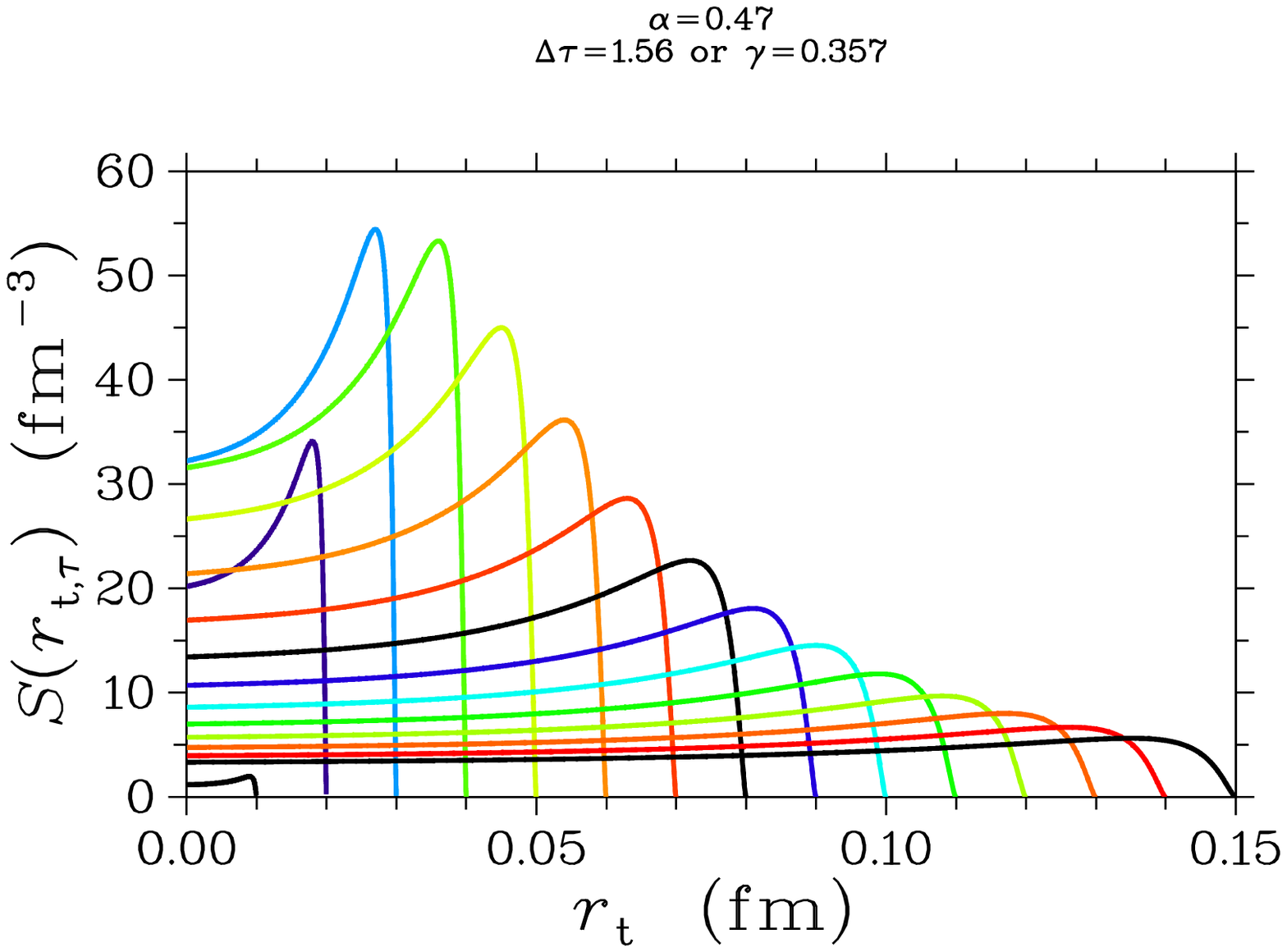}
  \caption{The transverse source function $S(x,y,\tau)$, \Eq{eq:Sxytau}, normalized to
           unity, and its transverse profile
           for various proper times ranging from 0.01\,fm to 0.15\,fm.
\label{fig:xymovie}
}
\end{figure}

\Fig{fig:xymovie} shows the transverse part of the emission function, \Eq{eq:Sxytau}, for
various proper times. Particle production starts immediately, increases rapidly and decreases slowly.
A ring-like structure, similar to the expanding, ring-like wave created by a pebble in a pond, is observed.
These pictures together form a movie representing this temporal process, which takes place,
to our knowledge, at the shortest time scale ever measured.\footnote{Animated gif files
covering the first 0.15\,fm ($0.5\times10^{-24}$\,s)
are available~\cite{tamas_movie}.}
In the movie two       ring-like structures separate along the thrust axis, their radii increasing.
They thus span the surface of two back-to-back cones whose tips meet at the origin.  The rings are
very faint at the beginning, then quickly brighten as the particle production probability increases to a
sharp peak at $\tau\approx0.035$\,fm ($\approx0.12\times10^{-24}$\,s).
These separating and expanding       rings then start to fade gradually
with a characteristic duration of $\Delta\tau\approx1.56$\,fm ($\approx5.2\times10^{-24}$s) following a power-law tail.
        Note however that the shape of these rings is in marked contrast
        to the more smoothly edged rings of fire found in hadron-proton reactions~\cite{NA22emiss}.
        In our case, these rings are reconstructed from
        the Bose-Einstein correlation functions and from the measured
        \pt\ and rapidity spectra of two-jet events without any reference to temperature or
        flow, indicating the non-thermal nature of particle production
        in \Pep\Pem\ annihilation.
 
\section{Test of dependence of BEC on components of \boldmath{$Q$}}            \label{sect:elongation}
 
In this section we return to the question of a possible elongation of the region of homogeneity 
and the possible dependence of the Bose-Einstein correlation function on components of $Q$.
 
The \taumodel\ predicts that the two-particle BEC correlation function $R_2$ depends on the two-particle
momentum difference only through $Q$, \ie, that a single radius parameter applies to all components of $Q$.
However,
previous studies \cite{L3_3D:1999,OPAL3D:2000,DELPHI2D:2000,ALEPH:2004,OPAL:2007} have found
that different radii are required for the different components of $Q$ in order to fit the data,
and indeed that the shape of the region of homogeneity is elongated along the event (thrust) axis.
The question is whether this is
an artifact of the Edgeworth or Gaussian parametrizations used in these studies
or shows a limitation of the \taumodel.
 
\subsection{Test of elongation and the \boldtaumodel}                          \label{subsect:elongation}
 
In the previous studies of elongation, $Q^2$ is split into three components,
\begin{align}   \label{eq-QLCMS}
  \Qsquare &=  \Qslong + \Qsside + \Qsout - (\Delta E)^2   \\
           &=  \Qslong + \Qsside + \Qsout \left(1 - \beta^2\right)  \;,  \qquad
 \text{where}\quad \beta   = \frac{p_{1\mathrm{out}}+p_{2\mathrm{out}}}{E_1+E_2}
\end{align}
in the Longitudinal Center of Mass System (LCMS) of the pair.
The LCMS frame is defined as the frame, obtained by a Lorentz boost along the event axis,
where the sum of the three-momenta of the two pions ($\vec{p}_1+\vec{p}_2$)
is perpendicular to the event axis.
Assuming azimuthal symmetry about the event axis suggests that the region of
homogeneity have an ellipsoidal  shape with one axis, the longitudinal axis, along the event axis.
The longitudinal axis would then be longer, shorter, or equal to the other two (transverse) axes, which are of equal length.
 
In the LCMS frame the event axis is referred to as the longitudinal direction;
the direction of $\vec{p}_1+\vec{p}_2$ as the out direction; and
the direction perpendicular to these as the side direction.
The components of $\abs{\vec{p}_1-\vec{p}_2}$ along these directions are denoted by
\Qlong, \Qout, and \Qside.
In the Gaussian and Edgeworth parametrizations $\Rsquare\Qsquare$ is then replaced in the fitting function by
\begin{equation}   \label{eq-RQLCMS}
    \Rsquare\Qsquare  \Longrightarrow \Rslong\Qslong + \Rsside\Qsside + \rhosout\Qsout \;.
\end{equation}
The longitudinal and transverse sizes of the source are measured by \Rlong\ and \Rside, respectively,
whereas the value of \rhoout\ reflects both the transverse and temporal sizes.\footnote{In the
literature the coefficient of \Qsout\ in \Eq{eq-RQLCMS} is usually denoted \Rsout. We prefer to use
\rhosout\ to emphasize that, unlike \Rlong\ and \Rside, \rhoout\ contains a dependence on $\beta$
and to differentiate it from \Rout\ in \Eq{eq-asymlevR2BPaLEB} below.}
Note that the LCMS and the rest frame of the pair differ only by a Lorentz boost of velocity $\beta$ along
the out direction.
Hence \Rlong\ and \Rside\ also measure the longitudinal and transverse size in the pair rest frame,
while \rhoout\ may be interpreted as
an average of the transverse size in the rest frame boosted to the LCMS:
$\rhosout=\rsout\langle 1-\beta^2\rangle$.
 
In our previous analysis~\cite{L3_3D:1999} we found, using all events,  $\Rside/\Rlong=0.80\pm0.02^{+0.03}_{-0.18}$
and $\Rside/\Rlong=0.81\pm0.02^{+0.03}_{-0.19}$ for the Gaussian and Edgeworth parametrizations, respectively.
Since the present analysis uses a somewhat different event sample and mixing algorithm, we have repeated these fits
using the same binning of $0.08\GeV$ for $0.00<Q_i<1.04\GeV$ ($i$=L, side, out).
We find consistent values~\cite{tamas:thesis} of $\Rside/\Rlong$: 0.76 for both parametrizations.
We also find the elongation to be greater for two-jet events than for three-jet events, as has previously been
observed by \OPAL\cite{OPAL3D:2000}.
Using the Edgeworth parametrization the elongation is
(statistical uncertainties only)
$\Rside/\Rlong = 0.65\pm0.02$ for two-jet events,
                $0.76\pm0.02$ for all events,
and
                $0.84\pm0.02$ for three-jet events.

We next investigate whether the \taumodel\ parametrizations could accommodate an elongation.
This is done by incorporating \Eq{eq-RQLCMS}
in \Eq{eq:asymlevR2}.  If \Eq{eq:asymlevRaR} is imposed, this results in
\begin{equation}\label{eq-asymlevR2BP}
    R_2(Q) = \gamma \left[ 1+ \lambda
                       \cos\left(\!\tan\left(\frac{\alpha\pi}{2}\right) A^{2\alpha} \vphantom{\frac{\Gamma}{2}}
                          \right)
             \exp \left(-A^{2\alpha} \right) \right]
             \left(1+ \epsillong\Qlong + \epsilside\Qside + \epsilout\Qout\right) \;,
\end{equation}
where
\begin{equation}\label{eq-asymlevR2BPA}
        A^2 = \Rslong\Qslong + \Rsside\Qsside + \rhosout\Qsout  \;,
\end{equation}
and where we use the same long-range parametrization as in the above Edgeworth and Gaussian parametrizations.
We have attempted the long-range parametrization $(1+\epsilon Q)$, but that results in unacceptable \chisq.
To extend the range of $Q$ included in the fit, the bin size is increased to $0.16\GeV$ for $Q_i>0.88\,\GeV$.
Results for two-jet events of a fit of \Eq{eq-asymlevR2BP} for $Q<4\,\GeV$
are shown in \Tab{tab:BPlcms}.  A clear preference for elongation ($\Rside/\Rlong<1$) is seen.
The value of $R$ found in the corresponding fit of $R_2(Q)$ (\Tab{tab:a_levy_c})
lies between the values of $\Rside$ and $\Rlong$ found here.
Furthermore, the value of the elongation agrees well with that found above for two-jet events using the Edgeworth parametrization.
 
\begin{table}
\caption{Results
         for two-jet events
         of fits
         of \Eq{eq-asymlevR2BP}.
         Uncertainties are statistical only.
\label{tab:BPlcms}
         }
\begin{center}
$
\begin{array}{ccc}
\hline
    \lambda                       &    0.49 \pm 0.02  \\ 
    \alpha                        &    0.46 \pm 0.01  \\ 
    \Rlong\text{ (fm)}            &    0.85 \pm 0.04  \\ 
    \Rside/\Rlong                 &    0.61 \pm 0.02  \\ 
    \rhoout/\Rlong                &    0.66 \pm 0.02  \\ 
    \epsillong\text{ (\invGeV)}   &    0.001\pm 0.001 \\ 
    \epsilside\text{ (\invGeV)}   &   -0.076\pm 0.003 \\ 
    \epsilout\text{ (\invGeV)}    &   -0.029\pm 0.002 \\ 
    \gamma                        &    1.011\pm 0.002 \\ 
\hline
    \chisq/\text{DoF}             &   14847/14921     \\ 
    \text{CL}                     &   \text{66\%}     \\ 
\hline
\end{array}
$
\end{center}
\end{table}
 
\subsection{Direct test of \boldmath{$R_2$} dependence on components of \boldmath{$Q$}}   \label{subsect:direct_test}
 
To directly test the hypothesis of no separate dependence on components of $Q$, we investigate
two decompositions of $Q$:
\begin{subequations}
\begin{align}               
  \Qsquare &=  \Qsle   + \Qsside + \Qsout   \label{eq-Q2dec-el}   \\
  \Qsquare &=  \Qslong + \Qsside + \qsout   \label{eq-Q2dec-q}
\end{align}
\end{subequations}
where $\Qsle=\Qslong - (\Delta E)^2$ and
      $\qsout=\Qsout - (\Delta E)^2$.
The decomposition of \Eq{eq-Q2dec-el} corresponds to the LCMS frame where the longitudinal and energy terms
are combined; its three components of $Q$ are invariant with respect to Lorentz boosts along the thrust axis.
The decomposition of \Eq{eq-Q2dec-q} corresponds to the LCMS frame boosted to the rest frame of the pair;
its three components of $Q$ are invariant with respect to Lorentz boosts along the out direction.
 
The test then consists of replacing \Rsquare\Qsquare\ by \Rsquare\ times one of the above equations and
comparing a fit where the coefficients of all three terms are constrained to be equal to a fit where each of
these coefficients is a free parameter.
 
Modifying \Eq{eq:asymlevR2} in this way, and imposing \Eq{eq:asymlevRaR}, results in
\begin{equation}\label{eq-asymlevR2BPm}
    R_2(Q) = \gamma \left[ 1+ \lambda
                       \cos\left(\!\tan\left(\frac{\alpha\pi}{2}\right) B^{2\alpha} \vphantom{\frac{\Gamma}{2}}
                          \right)
             \exp \left(-B^{2\alpha} \right) \right]
             b \;,
\end{equation}
where
\begin{subequations}   \label{eq-asymlevR2BPmLE}
\begin{align}
        B^2 &= \Rsle\Qsle + \Rsside\Qsside + \Rsout\Qsout                         \label{eq-asymlevR2BPaLEB}
\\
        b   &= 1+ \epsille\Qle + \epsilside\Qside + \epsilout\Qout                \label{eq-asymlevR2BPaLEb}
\end{align}
\end{subequations}
or
\begin{subequations}   \label{eq-asymlevR2BPmlsr}
\begin{align}
        B^2 &= \Rslong\Qslong + \Rsside\Qsside + \rsout\qsout               \\ 
        b   &= 1+ \epsillong\Qlong + \epsilside\Qside + \epsilout\qout  \;.    
\end{align}
\end{subequations}
 
Note that, since $\Delta E$ vanishes in the rest frame of the  pair, the quantity \rout\ is a direct measure of the spatial
extent of the source in this frame whereas \rhoout\ in \Eq{eq-asymlevR2BPA}
and \Rle\ in \Eq{eq-asymlevR2BPaLEB}  are also sensitive to the temporal distribution of the source.

The results of the four fits of \Eq{eq-asymlevR2BPm}
are shown in \Tab{tab:2jetBPRf}
for $Q<4\,\GeV$.
For both parametrizations the fit allowing separate dependence on the components of $Q$ has a higher
confidence level than the fit which does not.
The fit using  \Eq{eq-asymlevR2BPmLE} attains a confidence level of 2\% when no separate dependence is allowed.
While this is not poor enough to reject by itself the hypothesis of no elongation, the fit allowing separate dependence
describes the data better with a confidence level of 38\%.
When \Eq{eq-asymlevR2BPmlsr} is used the fit not allowing separate dependence
is rejected by a confidence level of $10^{-7}$, while allowing it results in the
acceptable confidence level of 2\%.
For the fits of both \Eq{eq-asymlevR2BPmLE} and \Eq{eq-asymlevR2BPmlsr} the differences in \chisq\ between
the fit allowing separate dependence and the fit not allowing it is huge, 296 and 464, respectively, while a difference of only 2
is expected if the hypothesis of no separate dependence is correct.
This provides extremely strong evidence against the hypothesis of identical dependence of $R_2$ on the
different components of $Q$.
 
\begin{figure}
  \centering
  \includegraphics[width=.94\textwidth,bb=58 87 519 682,clip]{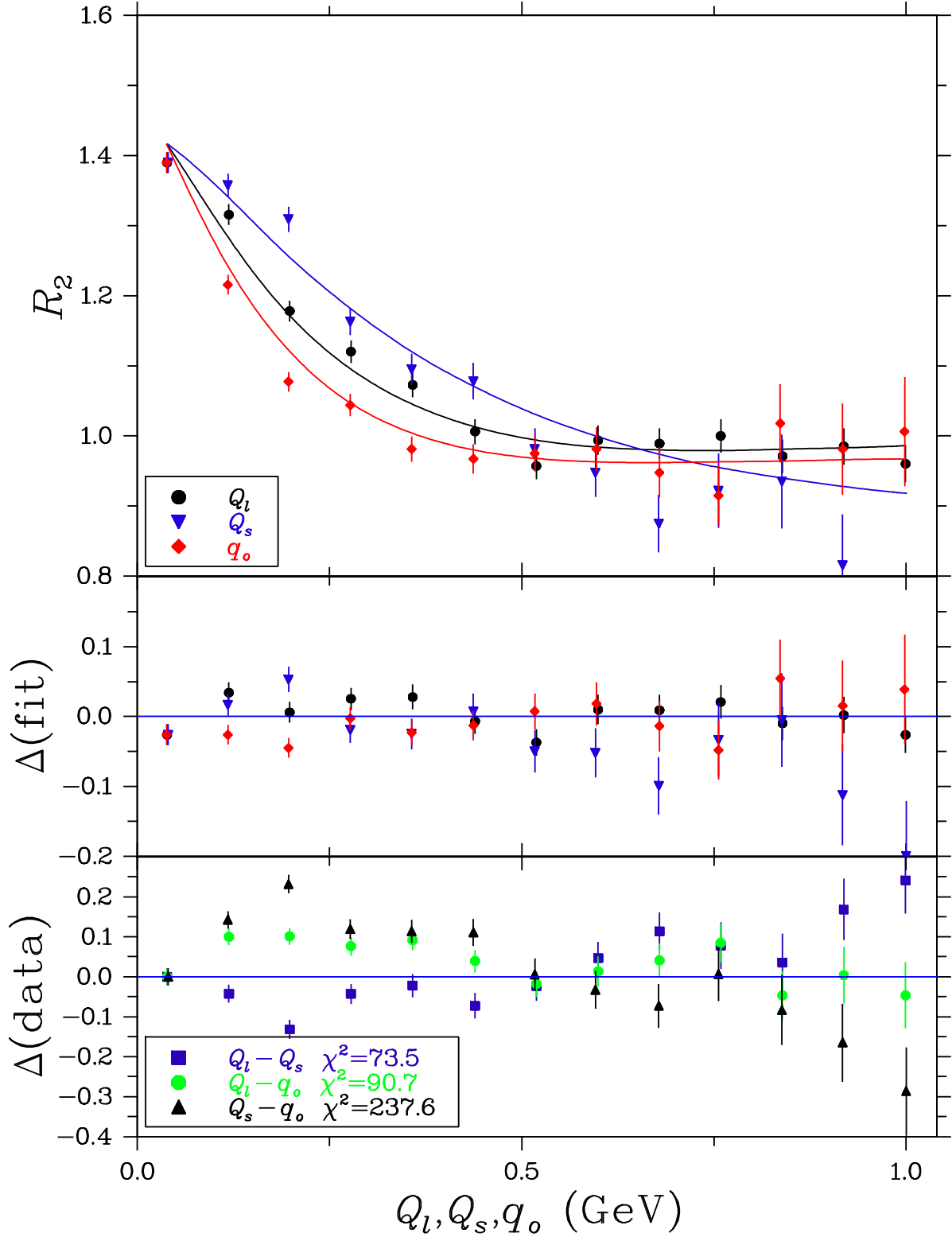}
  \caption{Projections of the Bose-Einstein correlation function $R_2(\Qlong,\Qside,\qout)$ for two-jet events
           using the region below 80 \MeV\ for the non-projected components.
           \label{fig-2jetBPRfr4}
           }
\end{figure}

The fits using \Eq{eq-asymlevR2BPmlsr}
are compared to the data in \Fig{fig-2jetBPRfr4} for small values of $Q$.
As is apparent in \Fig{fig-2jetBPRfr4}
the shape of the dependence of $R_2$ on the different components of $Q$
is different at small values of these components and can not be described by equal values of \Rlong, \Rside\ and \rout.
 
We have repeated the fits varying the upper limit of $Q$ to as low as 1\,\GeV.
While some parameters show a considerable dependence on the $Q$-range of the fit, $\Rside/\Rlong$ and $\rout/\Rlong$ are found to be
stable, varying by less than a standard deviation.
The variation in the other parameters  is thought to arise from correlations among the parameter values
and the inability to determine well the
values of the long-range parameters $\epsillong$,  $\epsilside$ and $\epsilout$ when $Q$ is confined to small values.
The value of the elongation, $\Rside/\Rlong$, is far from unity and agrees well with the value in the fit of \Eq{eq-asymlevR2BP}
imposing \Eq{eq:asymlevRaR} (\Tab{tab:BPlcms}) and with that found using the Edgeworth parametrization.
The value of $\rout/\Rlong$ is also far from unity, but is greater than unity, in contrast to $\Rside/\Rlong$.
Even more striking is the inequality of \Rside\ and \rout: $\rout/\Rside\simeq2$. This indicates the invalidity of
the azimuthal symmetry assumed in the elongation analyses.

\begin{table}
\caption{Results
         for two-jet events
         of fits
         of \Eq{eq-asymlevR2BPm} and \Eq{eq-asymlevR2BPmLE} or (\ref{eq-asymlevR2BPmlsr}) for $Q<4\,\GeV$
         with all radii as free parameters and with them constrained to be equal.
         Uncertainties are statistical only.
\label{tab:2jetBPRf}
         }
\begin{center}
$
\begin{array}{rccc}
\hline  \text{\Eq{eq-asymlevR2BPmLE}}
 & \lambda                      &    0.51 \pm 0.03   &   0.49 \pm 0.03   \\
 & \alpha                       &    0.46 \pm 0.01   &   0.46 \pm 0.01   \\
 & \Rle\text{ (fm)}             &    0.84 \pm 0.04   &   0.71 \pm 0.04   \\
 & \Rside/\Rle                  &    0.60 \pm 0.02   &        1          \\
 & \Rout/\Rle                   &    0.986\pm 0.003  &        1          \\
 & \epsille\text{ (\invGeV)}    &    0.001\pm 0.001  &   0.000\pm 0.001  \\
 & \epsilside\text{ (\invGeV)}  &   -0.069\pm 0.003  &  -0.064\pm 0.003  \\
 & \epsilout\text{ (\invGeV)}   &   -0.032\pm 0.002  &  -0.035\pm 0.002  \\
 & \gamma                       &    1.010\pm 0.002  &   1.012\pm 0.002  \\
\hline
 &  \chisq/\text{DoF}           &     14590/14538    &    14886/14540    \\
 &  \text{CL}                   &     \text{38\%}    &    \text{2\%}     \\
   \hline
\hline  \text{\Eq{eq-asymlevR2BPmlsr}}
 &   \lambda                        &    0.65 \pm 0.03   &   0.57 \pm 0.03  \\
 &   \alpha                         &    0.41 \pm 0.01   &   0.44 \pm 0.01  \\
 &   \Rlong\text{ (fm)}             &    0.96 \pm 0.05   &   0.82 \pm 0.04  \\
 &   \Rside/\Rlong                  &    0.62 \pm 0.02   &        1         \\
 &   \rout/\Rlong                   &    1.23 \pm 0.03   &        1         \\
 &   \epsillong\text{ (\invGeV)}    &    0.004\pm 0.001  &   0.003\pm 0.001 \\
 &   \epsilside\text{ (\invGeV)}    &   -0.067\pm 0.003  &  -0.059\pm 0.003 \\
 &   \epsilout\text{ (\invGeV)}     &   -0.022\pm 0.003  &  -0.029\pm 0.002 \\
 &   \gamma                         &    1.000\pm 0.002  &   1.003\pm 0.002 \\
\hline
 &   \chisq/\text{DoF}              &     10966/10647    &    11430/10649   \\
 &   \text{CL}                      &     \text{2\%}     &    10^{-7}       \\
\hline
\end{array}
$
\end{center}
\end{table}

We also investigate elongation in the parametrization of \Eq{eq:levyR2a} by replacing $\Delta\tau Q^2$ by
$\Delta\tau_\text{long}\Qslong+\Delta\tau_\text{side}\Qsside+\Delta\tau_\text{out}\qsout$.
Fits of the resulting parametrization are performed both with
$\Delta\tau_\text{long}$, $\Delta\tau_\text{side}$ and $\Delta\tau_\text{out}$ as free parameters
and with them constrained to be equal.
In these fits only one \mt\ bin is used, $0<\mt<4\GeV$.
The results are shown in
\Tab{tab:2jetBPmtf}.                              
%
The conclusions are the same as for the fits of \Eq{eq-asymlevR2BPm}.
The value of $\Deltatauside/\Deltataulong$ agrees with that of $\Rside/\Rlong$, and \Deltatauout\ is greater than unity.
Constraining the fit by $\Deltataulong=\Deltatauside=\Deltatauout$ results in an unacceptable \chisq.

We note that in all of the fits of \adhoc\/ \taumodel\ parametrizations the long-range correlation
parameters, \epsilside\ and \epsilout, deviate substantially and significantly from zero.
This could indicate that not all non-BEC correlations are removed from the reference sample, which in turn could
influence the amount of elongation found by the fits.
To investigate this we have performed fits of the region $Q>1.8\GeV$ fixing the BEC contribution to zero
($\lambda=0$).  The fits for $Q<4\GeV$ were then performed with the $\epsilon_i$ fixed to the values
obtained in the $\lambda=0$ fits.  Of course, the confidence levels were lower than those of the fits where
the $\epsilon_i$ could vary, but the elongation varied only slightly.  For the fits of \Tab{tab:2jetBPRf}
$\Rside/\Rle=0.58\pm0.02$ and $\Rout/\Rle=0.984\pm0.002$ with a CL of 8\%, while
$\Rside/\Rlong=0.64\pm0.02$ and $\Rout/\Rlong=1.25\pm0.03$ with a CL of 0.09\%.
We also repeated the fits with the $\epsilon_i$ fixed to zero.  The confidence levels were much worse,
but the elongation parameters proved quite robust:
$\Rside/\Rle=0.69\pm0.02$ and $\Rout/\Rle=0.99\pm0.04$ with a CL of $10^{-11}$, and
$\Rside/\Rlong=0.72\pm0.02$ and $\Rout/\Rlong=1.24\pm0.03$ with a CL of $10^{-13}$.
Further, note in \Fig{fig-2jetBPRfr4} that the data themselves show different dependencies on \Qlong, \Qside, \qout.
We conclude that the elongation is real, \ie, not an artifact of the parametrizations previously used,
and that the \adhoc\/ modified \taumodel\ provides a reasonable description of the elongation.

\begin{table}
\caption{Results
         for two-jet events
         of fits
         of \Eq{eq:levyR2a} with $\Delta\tau Q^2$ replaced by
         $\Delta\tau_\text{long}\Qslong+\Delta\tau_\text{side}\Qsside+\Delta\tau_\text{out}\qsout$
         and long-range parametrization     $1+ \epsillong\Qlong + \epsilside\Qside + \epsilout\qout$,
         where $\epsillong$, $\epsilside$ and $\epsilout$  are free parameters.
         In the first fit all three $\Delta\tau$ parameters are free; in the second they are constrained to be equal.
         Uncertainties are statistical only.
\label{tab:2jetBPmtf}
         }
\begin{center}
$
\begin{array}{ccc}
\hline
    \lambda                        &  0.57 \pm0.03  &  0.49 \pm0.02 \\
    \alpha                         &  0.47 \pm0.01  &  0.50 \pm0.01 \\
    \Deltataulong\text{ (fm)}      &  1.74 \pm0.15  &  1.32 \pm0.10 \\
    \Deltatauside/\Deltataulong    &  0.61 \pm0.03  &       1       \\
    \Deltatauout/\Deltataulong     &  1.48 \pm0.08  &       1       \\
    \epsillong\text{ (\invGeV)}    &  0.000\pm0.001 & -0.001\pm0.001\\
    \epsilside\text{ (\invGeV)}    & -0.064\pm0.003 & -0.056\pm0.002\\
    \epsilout\text{ (\invGeV)}     & -0.021\pm0.003 & -0.025\pm0.002\\
    \gamma                         &  1.008\pm0.002 &  1.010\pm0.002\\
\hline
    \chisq/\text{DoF}              &   10977/10647  &   11239/10649 \\
    \text{CL}                      &   \text{1\%}   &   10^{-5}     \\
\hline
\end{array}
$
\end{center}
\end{table}

\section{Discussion and Conclusions}  \label{sect:discussion}
The usual parametrizations of BEC of pion pairs in terms of their four-momentum difference $Q$,
such as the Gaussian,  Edgeworth, or symmetric L\'evy parametrizations, are found to be incapable
of describing the Bose-Einstein correlation function $R_2$, particularly the  anti-correlation region,
approximately $0.5<Q<1.5$\,\GeV.
This failure has not been previously so obvious, since previous analyses generally fit $R_2(Q)$ only
up to 2\,\GeV\ or less and the anti-correlation region is then tacitly absorbed into the
parametrization of long-range correlations.
The existence of this anti-correlation region has the more general implication of ruling out the
form usually proposed for BEC, $R_2=1+\lvert\tilde{f}(Q)\rvert^2$.

Those parametrizations of BEC are based on a static view of pion emission.  No time dependence
is assumed, which means that either the pion emission volume is unchanging during pion emission or that the
parameters describing the volume, which result from fitting $R_2$, are some sort of time average.
 
The \taumodel\ assumes a very high degree of correlation between momentum space and space-time
and introduces a time dependence explicitly.
The Bose-Einstein correlation function $R_2$ measures the real part of
products of Fourier-transformed proper-time distributions. Not only the positive
correlations at low values $Q$ but also negative correlations at somewhat higher values of $Q$ are
predicted.  It is as though the Bose-Einstein symmetry pulls identical pions below some critical value of
$Q$ closer
together creating an excess of pairs at lower $Q$ and leaving a deficit of pairs around this value of $Q$.
 
In the \taumodel\ $R_2$ is then a function of one two-particle variable,
the invariant four-momentum difference of the two particles $Q$, and of two single-particle variables,
the values of $a$ of the two particles, where
$a$ is a parameter in the correlation between momentum space and space-time,  \Eq{eq:R2cor},
The time-dependence is assumed to be given by an asymmetric L\'evy distribution, $H(\tau)$,
which imposes causality by allowing particle production only for $\tau>0$.
The parameter $a$ can be absorbed into an effective radius to obtain an expression, \Eq{eq:asymlevR2}, for $R_2(Q)$ which
successfully describes BEC, including the anti-correlation region, for both two- and three-jet events.
We note that a similar anti-correlation region has recently been observed by the CMS Collaboration~\cite{CMS:be2} in pp collisions at
$\sqrt{s}=0.9$ and 7\,\TeV\ and successfully fit by the \taumodel\ formula,  \Eq{eq:asymlevR2}.

For two-jet events $a=1/\mt$ and the introduction of an effective radius is unnecessary.  The BEC
correlation
function is then a function not only of $Q$ but also of the transverse masses of the pions.  This
description of $R_2(Q,{\mt}_1,{\mt}_2)$ also successfully describes the data.
 
Nevertheless, the \taumodel\ description breaks down when confronted with data expressed in components
of $Q$.
The \taumodel\  predicts   that the only two-particle variable entering $R_2$ is $Q$, \ie,
that there is a single radius parameter which is the same for each of the separate components of $Q$.
This implies a spherical shape of the pion
region of homogeneity, 
whereas previous analyses using static parametrizations have found a shape elongated along the event axis,
as evidenced by \Rside\ being smaller than \Rlong\ in the LCMS.  Assuming azimuthal symmetry about the event axis,
\Rside\ is the transverse radius of the region of homogeneity.
 
Accordingly, the \taumodel\ equations for $R_2$ have been modified \adhoc\/ to allow an elongation.
Moreover, fits have been performed not only in the LCMS, but also in the rest frame of the pair.
Fits not allowing elongation have much worse confidence levels than fits allowing elongation.
The elongation found
agrees with that found in the previous analyses. 
However, in the rest frame of the pair the radius parameter in the out direction is found to be approximately twice that in
the side direction. Thus the assumption of azimuthal symmetry about the event axis is found not to hold.
 
We note that an elongation is expected~\cite{AR98a} for a hadronizing string in the Lund model,
where the transverse and longitudinal components of a particle's momentum arise from different mechanisms.
Perhaps the absence of azimuthal symmetry arises from gluon radiation, which is represented by kinks in the Lund string.
Therefore a possible improvement of the \taumodel\ could be to allow the longitudinal and transverse components
to have  different degrees of
correlation between a particle's momentum and its space-time production point in \Eq{eq:tau-corr}.

The \taumodel, as we have used it, incorporates a L\'evy distribution.
A L\'evy distribution arises naturally from a fractal,
or from a random walk or anomalous diffusion~\cite{metzler},
and the parton shower of the leading log
approximation of QCD is a fractal \cite{Dahlqvist:1989yc,Gustafson:1990qi,Gustafson:1991ru}.
In this case,
the L\'evy index of stability is related to the strong coupling
constant, \alphas, by\cite{alphasLevy,alphasLevy:arx}
\begin{equation}
    \alphas = \frac{2\pi}{3} \alpha^2 \;.
\end{equation}
Assuming (generalized) local parton hadron duality \cite{Azimov:lphd,Azimov:lphd1,genlphd},
one can expect that the distribution of hadrons retains the features of the gluon distribution.
For the value of $\alpha$ found in the fit of \Eq{eq:levyR2a} 
for two-jet events (\Tab{tab:2jetR2}) we find
$\alphas=0.46\pm0.01^{+0.04}_{-0.02}$.
This is a reasonable value for a scale of the order of 1\,\GeV,  
which is the scale at which the production of hadrons is thought to take place.
For comparison, from $\tau$ decay, $\alphas(m_\tau\approx1.8\GeV)=0.34\pm0.03$ \cite{PDG:2008},
and from the average value of $\alphas(M_\mathrm{Z})$,  $\alphas(1\GeV)=0.50^{+0.06}_{-0.05}$ \cite{runningalpha}.

It is of particular interest to point out the \mt\ dependence of the ``width'' of the source.
In \Eq{eq:levyR2a}
the parameter associated with the width is $\Delta \tau$.
Note that it enters \Eq{eq:levyR2a} 
as $\Delta\tau Q^2/\overline{m}_\mathrm{t}$.
In a Gaussian parametrization the radius $R$ enters the parametrization as $R^2Q^2$.
Our observation that $\Delta\tau$ is independent of $\overline{m}_\mathrm{t}$ thus corresponds to
$R\propto1/\sqrt{\overline{m}_\mathrm{t}}$
and can be interpreted as a confirmation of the observation~\cite{Smirnova:Nijm96,Dalen:Maha98,OPAL:2007}
of such a dependence of the Gaussian radii in 2- and 3-dimensional analyses of Z decays.
The lack of  dependence of all the parameters of \Eq{eq:levyR2a} 
on the transverse mass is in
accordance with the \taumodel.

Given these successes,
the BEC fit results of the \taumodel\ have been used to reconstruct
the pion emission function of two-jet events.
Note that all charged tracks are treated as pions.
For BEC the main influence of this is a lowering of the value of the parameter $\lambda$.
However, the influence of this as well as the influence of resonances
on the rapidity and \pt\ distributions may be more profound.
Hence, the emission function is subject to some quantitative uncertainty.
Nevertheless, the general features of the emission function should remain valid:
Particle production begins immediately after collision, increases rapidly and then decreases slowly,
occuring predominantly close to the light cone.
In the transverse plane a ring-like structure expands outwards,
which is similar to the picture in hadron-hadron interactions
but unlike that of heavy-ion collisions~\cite{Tamas:HIP2002}.
Despite this similarity the physical process is much different.
Reflecting the non-thermal nature of \Pep\Pem\ annihilation, the proper-time distribution
and the space-time structure are reconstructed here without any reference to a temperature, in
contrast to the results of earlier hadron-hadron and heavy-ion collisions.
 
%
\section*{Acknowledgments}
We are grateful to J. P. Nolan for an academic licence for his Mathematica package on Levy Stable
distributions.
 
%
%
\section*{Appendix}
In \Tabs{tab:2jetdata} and \ref{tab:3jetdata} we present the data used in the
fits of Section~\ref{sect:param}.
 
\begin{center}
\begin{longtable}{cccr@{$\pm$}l}
\caption{Data for fits to $R_2(Q)$
         for two-jet events
         used in \Figs{fig:gauss_2jet}, \ref{fig:a_levy_2jet}, and \ref{fig:2jetR2},
         as well as
         \Tabs{tab:a_levy}, \ref{tab:a_levy_c}, and \ref{tab:2jetR2}.
         Uncertainties are statistical only.
         }
\label{tab:2jetdata}
\\
 $ Q\:\text{(\GeV)}$ & $\langle{\mt}_1\rangle\:\text{(\GeV)}$  & $\langle{\mt}_2\rangle\:\text{(\GeV)}$   & \multicolumn{2}{c}{$R_2$} \\
 \hline
\endfirsthead
\multicolumn{5}{c}%
{{\bfseries \tablename\ \thetable{} -- continued from previous page}}  \\ \hline
 $ Q\:\text{(\GeV)}$ & $\langle{\mt}_1\rangle\:\text{(\GeV)}$  & $\langle{\mt}_2\rangle\:\text{(\GeV)}$   & \multicolumn{2}{c}{$R_2$} \\
\hline
\endhead
 
\hline \multicolumn{5}{r}{{Continued on next page}} \\ \hline
\endfoot
\\
\endlastfoot
%
 0.030 & 0.265 & 0.245 &  1.362 & 0.051  \\
 0.064 & 0.283 & 0.238 &  1.424 & 0.021  \\
 0.102 & 0.303 & 0.232 &  1.362 & 0.013  \\
 0.141 & 0.320 & 0.225 &  1.270 & 0.009  \\
 0.181 & 0.337 & 0.222 &  1.210 & 0.007  \\
 0.221 & 0.354 & 0.219 &  1.154 & 0.006  \\
 0.260 & 0.370 & 0.218 &  1.118 & 0.005  \\
 0.300 & 0.387 & 0.220 &  1.091 & 0.005  \\
 0.340 & 0.402 & 0.222 &  1.062 & 0.005  \\
 0.380 & 0.418 & 0.225 &  1.028 & 0.004  \\
 0.420 & 0.432 & 0.229 &  1.023 & 0.004  \\
 0.460 & 0.445 & 0.233 &  1.005 & 0.004  \\
 0.500 & 0.457 & 0.238 &  0.991 & 0.004  \\
 0.540 & 0.472 & 0.243 &  0.979 & 0.004  \\
 0.580 & 0.485 & 0.247 &  0.968 & 0.004  \\
 0.620 & 0.496 & 0.252 &  0.965 & 0.004  \\
 0.660 & 0.508 & 0.256 &  0.962 & 0.004  \\
 0.700 & 0.519 & 0.261 &  0.959 & 0.004  \\
 0.740 & 0.531 & 0.265 &  0.959 & 0.004  \\
 0.780 & 0.540 & 0.268 &  0.959 & 0.004  \\
 0.820 & 0.549 & 0.271 &  0.958 & 0.004  \\
 0.860 & 0.559 & 0.275 &  0.957 & 0.005  \\
 0.900 & 0.565 & 0.278 &  0.963 & 0.005  \\
 0.940 & 0.572 & 0.281 &  0.958 & 0.005  \\
 0.980 & 0.579 & 0.284 &  0.965 & 0.005  \\
 1.020 & 0.582 & 0.285 &  0.962 & 0.005  \\
 1.060 & 0.589 & 0.287 &  0.967 & 0.005  \\
 1.100 & 0.594 & 0.290 &  0.965 & 0.005  \\
 1.140 & 0.599 & 0.292 &  0.962 & 0.005  \\
 1.180 & 0.602 & 0.293 &  0.969 & 0.005  \\
 1.220 & 0.605 & 0.295 &  0.962 & 0.005  \\
 1.260 & 0.611 & 0.296 &  0.988 & 0.006  \\
 1.300 & 0.612 & 0.298 &  0.982 & 0.006  \\
 1.340 & 0.617 & 0.298 &  0.985 & 0.006  \\
 1.380 & 0.616 & 0.300 &  0.979 & 0.006  \\
 1.420 & 0.619 & 0.301 &  0.979 & 0.006  \\
 1.460 & 0.621 & 0.301 &  0.992 & 0.006  \\
 1.500 & 0.625 & 0.302 &  0.975 & 0.006  \\
 1.540 & 0.622 & 0.303 &  0.988 & 0.007  \\
 1.580 & 0.623 & 0.304 &  0.990 & 0.007  \\
 1.620 & 0.629 & 0.305 &  0.983 & 0.007  \\
 1.660 & 0.627 & 0.304 &  0.991 & 0.007  \\
 1.700 & 0.630 & 0.306 &  0.995 & 0.007  \\
 1.740 & 0.631 & 0.307 &  0.984 & 0.007  \\
 1.780 & 0.629 & 0.306 &  0.990 & 0.007  \\
 1.820 & 0.636 & 0.308 &  0.982 & 0.007  \\
 1.860 & 0.632 & 0.307 &  0.993 & 0.007  \\
 1.900 & 0.635 & 0.309 &  0.979 & 0.007  \\
 1.940 & 0.634 & 0.308 &  0.986 & 0.008  \\
 1.980 & 0.635 & 0.309 &  0.991 & 0.008  \\
 2.020 & 0.634 & 0.311 &  0.995 & 0.008  \\
 2.060 & 0.631 & 0.309 &  0.998 & 0.008  \\
 2.100 & 0.636 & 0.310 &  0.993 & 0.008  \\
 2.140 & 0.634 & 0.310 &  0.990 & 0.008  \\
 2.180 & 0.637 & 0.311 &  0.998 & 0.008  \\
 2.220 & 0.636 & 0.312 &  0.985 & 0.008  \\
 2.260 & 0.638 & 0.312 &  1.000 & 0.009  \\
 2.300 & 0.635 & 0.311 &  0.975 & 0.008  \\
 2.340 & 0.635 & 0.312 &  0.998 & 0.009  \\
 2.380 & 0.636 & 0.313 &  0.998 & 0.009  \\
 2.420 & 0.634 & 0.312 &  0.978 & 0.009  \\
 2.460 & 0.636 & 0.311 &  0.986 & 0.009  \\
 2.500 & 0.637 & 0.314 &  0.985 & 0.009  \\
 2.540 & 0.635 & 0.312 &  1.000 & 0.009  \\
 2.580 & 0.636 & 0.312 &  0.989 & 0.009  \\
 2.620 & 0.635 & 0.313 &  0.988 & 0.009  \\
 2.660 & 0.633 & 0.312 &  0.999 & 0.010  \\
 2.700 & 0.635 & 0.311 &  0.981 & 0.010  \\
 2.740 & 0.635 & 0.313 &  0.983 & 0.010  \\
 2.780 & 0.633 & 0.311 &  1.001 & 0.010  \\
 2.820 & 0.632 & 0.312 &  1.003 & 0.010  \\
 2.860 & 0.633 & 0.312 &  1.006 & 0.010  \\
 2.900 & 0.633 & 0.313 &  0.981 & 0.010  \\
 2.940 & 0.631 & 0.310 &  1.010 & 0.011  \\
 2.980 & 0.632 & 0.311 &  1.018 & 0.011  \\
 3.020 & 0.629 & 0.310 &  0.993 & 0.011  \\
 3.060 & 0.634 & 0.312 &  0.990 & 0.011  \\
 3.100 & 0.631 & 0.310 &  1.005 & 0.011  \\
 3.140 & 0.635 & 0.310 &  0.983 & 0.011  \\
 3.180 & 0.634 & 0.313 &  0.991 & 0.011  \\
 3.220 & 0.635 & 0.311 &  0.998 & 0.011  \\
 3.260 & 0.626 & 0.309 &  1.001 & 0.011  \\
 3.300 & 0.633 & 0.312 &  0.994 & 0.011  \\
 3.340 & 0.634 & 0.312 &  0.981 & 0.011  \\
 3.380 & 0.633 & 0.310 &  0.980 & 0.011  \\
 3.420 & 0.632 & 0.311 &  0.983 & 0.012  \\
 3.460 & 0.630 & 0.310 &  0.982 & 0.012  \\
 3.500 & 0.632 & 0.310 &  0.989 & 0.012  \\
 3.540 & 0.636 & 0.311 &  0.990 & 0.012  \\
 3.580 & 0.633 & 0.311 &  0.996 & 0.012  \\
 3.620 & 0.629 & 0.310 &  1.001 & 0.012  \\
 3.660 & 0.633 & 0.312 &  0.999 & 0.012  \\
 3.700 & 0.633 & 0.311 &  0.992 & 0.012  \\
 3.740 & 0.635 & 0.310 &  0.988 & 0.012  \\
 3.780 & 0.635 & 0.313 &  0.999 & 0.013  \\
 3.820 & 0.636 & 0.312 &  0.986 & 0.013  \\
 3.860 & 0.637 & 0.309 &  0.997 & 0.013  \\
 3.900 & 0.631 & 0.309 &  0.979 & 0.013  \\
 3.940 & 0.633 & 0.309 &  0.992 & 0.013  \\
 3.980 & 0.636 & 0.313 &  1.009 & 0.013  \\
\hline
\end{longtable}
\end{center}

\begin{table}
\caption{Data for fits to $R_2(Q)$
         for three-jet events
         used in \Figs{fig:gauss_3jet} and \ref{fig:a_levy_3jet}
         as well as
         \Tabs{tab:a_levy} and \ref{tab:a_levy_c}.
         Uncertainties are statistical only.
\label{tab:3jetdata}
         }
\begin{center}
$
\begin{array}{cr@{\pm}l@{\hspace{10mm}}cr@{\pm}l@{\hspace{10mm}}cr@{\pm}l}
\hline
   Q\:\text{(\GeV)}          & \multicolumn{2}{c}{R_2} &
   Q\:\text{(\GeV)}          & \multicolumn{2}{c}{R_2} &
   Q\:\text{(\GeV)}          & \multicolumn{2}{c}{R_2}  \\
\hline
  0.030 &    1.649 & 0.056 &  1.380 &    0.989 & 0.005  &  2.700  &   0.992 & 0.008 \\
  0.064 &    1.520 & 0.020 &  1.420 &    0.981 & 0.005  &  2.740  &   1.001 & 0.008 \\
  0.102 &    1.418 & 0.012 &  1.460 &    0.989 & 0.005  &  2.780  &   0.995 & 0.008 \\
  0.141 &    1.307 & 0.008 &  1.500 &    0.996 & 0.005  &  2.820  &   0.994 & 0.008 \\
  0.181 &    1.222 & 0.007 &  1.540 &    0.987 & 0.005  &  2.860  &   1.003 & 0.008 \\
  0.221 &    1.158 & 0.005 &  1.580 &    0.992 & 0.005  &  2.900  &   0.987 & 0.008 \\
  0.260 &    1.111 & 0.005 &  1.620 &    0.997 & 0.005  &  2.940  &   1.011 & 0.008 \\
  0.300 &    1.073 & 0.004 &  1.660  &   0.998 & 0.005  &  2.980  &   0.994 & 0.008 \\
  0.340 &    1.043 & 0.004 &  1.700  &   0.994 & 0.005  &  3.020  &   0.997 & 0.009 \\
  0.380 &    1.028 & 0.004 &  1.740  &   0.996 & 0.005  &  3.060  &   0.995 & 0.009 \\
  0.420 &    0.997 & 0.004 &  1.780  &   0.999 & 0.006  &  3.100  &   0.991 & 0.009 \\
  0.460 &    0.989 & 0.004 &  1.820  &   0.988 & 0.006  &  3.140  &   0.996 & 0.009 \\
  0.500 &    0.976 & 0.003 &  1.860  &   0.996 & 0.006  &  3.180  &   0.996 & 0.009 \\
  0.540 &    0.965 & 0.003 &  1.900  &   0.995 & 0.006  &  3.220  &   1.007 & 0.009 \\
  0.580 &    0.965 & 0.003 &  1.940  &   1.003 & 0.006  &  3.260  &   0.996 & 0.009 \\
  0.620 &    0.958 & 0.003 &  1.980  &   0.988 & 0.006  &  3.300  &   0.981 & 0.009 \\
  0.660 &    0.957 & 0.003 &  2.020  &   0.998 & 0.006  &  3.340  &   0.991 & 0.009 \\
  0.700 &    0.948 & 0.003 &  2.060  &   0.999 & 0.006  &  3.380  &   1.007 & 0.010 \\
  0.740 &    0.957 & 0.003 &  2.100  &   0.991 & 0.006  &  3.420  &   0.996 & 0.010 \\
  0.780 &    0.951 & 0.004 &  2.140  &   0.997 & 0.006  &  3.460  &   0.986 & 0.010 \\
  0.820 &    0.954 & 0.004 &  2.180  &   1.003 & 0.006  &  3.500  &   0.997 & 0.010 \\
  0.860 &    0.961 & 0.004 &  2.220  &   0.999 & 0.007  &  3.540  &   1.016 & 0.010 \\
  0.900 &    0.952 & 0.004 &  2.260  &   0.990 & 0.007  &  3.580  &   0.995 & 0.010 \\
  0.940 &    0.961 & 0.004 &  2.300  &   0.994 & 0.007  &  3.620  &   1.010 & 0.010 \\
  0.980 &    0.966 & 0.004 &  2.340  &   1.008 & 0.007  &  3.660  &   1.015 & 0.011 \\
  1.020 &    0.964 & 0.004 &  2.380  &   1.004 & 0.007  &  3.700  &   0.989 & 0.010 \\
  1.060 &    0.963 & 0.004 &  2.420  &   1.005 & 0.007  &  3.740  &   1.008 & 0.011 \\
  1.100 &    0.967 & 0.004 &  2.460  &   0.995 & 0.007  &  3.780  &   0.988 & 0.011 \\
  1.140 &    0.974 & 0.004 &  2.500  &   0.991 & 0.007  &  3.820  &   1.008 & 0.011 \\
  1.180 &    0.978 & 0.004 &  2.540  &   0.986 & 0.007  &  3.860  &   0.991 & 0.011 \\
  1.220 &    0.981 & 0.004 &  2.580  &   0.997 & 0.007  &  3.900  &   1.004 & 0.011 \\
  1.260 &    0.977 & 0.004 &  2.620  &   1.000 & 0.008  &  3.940  &   1.013 & 0.011 \\
  1.300 &    0.980 & 0.004 &  2.660  &   1.015 & 0.008  &  3.980  &   1.000 & 0.011 \\
  1.340 &    0.980 & 0.005 &  \multicolumn{6}{c}{\ }  \\
\hline
\end{array}
$
\end{center}
\end{table}

%
%
\newpage
\section*{Author List}
\typeout{   }
\typeout{Using author list for paper 287 -  }
\typeout{$Modified: Jul 15 2001 by smele $}
\typeout{!!!!  This should only be used with document option a4p!!!!}
\typeout{   }
%
%
%
%
%
%
 
\newcount\tutecount  \tutecount=0
\def\tutenum#1{\global\advance\tutecount by 1 \xdef#1{\the\tutecount}}
\def\tute#1{$^{#1}$}
\tutenum\aachen            
\tutenum\nikhef            
\tutenum\mich              
\tutenum\lapp              
\tutenum\basel             
\tutenum\lsu               
\tutenum\beijing           
\tutenum\bologna           
\tutenum\tata              
\tutenum\ne                
\tutenum\bucharest         
\tutenum\budapest          
\tutenum\mit               
\tutenum\panjab            
\tutenum\debrecen          
\tutenum\dublin            
\tutenum\florence          
\tutenum\cern              
\tutenum\wl                
\tutenum\geneva            
\tutenum\hamburg           
\tutenum\hefei             
\tutenum\lausanne          
\tutenum\lyon              
\tutenum\madrid            
\tutenum\florida           
\tutenum\milan             
\tutenum\moscow            
\tutenum\naples            
\tutenum\cyprus            
\tutenum\nymegen           
\tutenum\caltech           
\tutenum\perugia           
\tutenum\peters            
\tutenum\cmu               
\tutenum\potenza           
\tutenum\prince            
\tutenum\riverside         
\tutenum\rome              
\tutenum\salerno           
\tutenum\ucsd              
\tutenum\sofia             
\tutenum\korea             
\tutenum\taiwan            
\tutenum\tsinghua          
\tutenum\purdue            
\tutenum\psinst            
\tutenum\zeuthen           
\tutenum\eth               

{
\parskip=0pt
\noindent
{\bf The L3 Collaboration:}
\ifx\selectfont\undefined
 \baselineskip=10.8pt
 \baselineskip\baselinestretch\baselineskip
 \normalbaselineskip\baselineskip
 \ixpt
\else
 \fontsize{9}{10.8pt}\selectfont
\fi
\medskip
\tolerance=10000
\hbadness=5000
\raggedright
\hsize=162truemm\hoffset=0mm
\def\r{\rlap,}
\noindent
 
P.Achard\r\tute\geneva\
O.Adriani\r\tute{\florence}\
M.Aguilar-Benitez\r\tute\madrid\
J.Alcaraz\r\tute{\madrid}\
G.Alemanni\r\tute\lausanne\
J.Allaby\r\tute{\cern,\dagger}\
A.Aloisio\r\tute\naples\
M.G.Alviggi\r\tute\naples\
H.Anderhub\r\tute\eth\
V.P.Andreev\r\tute{\lsu,\peters}\
% ...
 
\rule{\textwidth}{0.4pt}
\begin{list}{A}{\itemsep=0pt plus 0pt minus 0pt\parsep=0pt plus 0pt minus 0pt
                \topsep=0pt plus 0pt minus 0pt}
\item[\aachen]
 III. Physikalisches Institut, RWTH, D-52056 Aachen, Germany$^{\S}$
\item[\nikhef] National Institute for High Energy Physics, NIKHEF,
     and University of Amsterdam, NL-1009 DB Amsterdam, The Netherlands
\item[\mich] University of Michigan, Ann Arbor, MI 48109, USA
\item[\lapp] Laboratoire d'Annecy-le-Vieux de Physique des Particules,
     LAPP,IN2P3-CNRS, BP 110, F-74941 Annecy-le-Vieux CEDEX, France
\item[\basel] Institute of Physics, University of Basel, CH-4056 Basel,
     Switzerland
\item[\lsu] Louisiana State University, Baton Rouge, LA 70803, USA
\item[\beijing] Institute of High Energy Physics, IHEP,
  100039 Beijing, China$^{\triangle}$
\item[\bologna] University of Bologna and INFN-Sezione di Bologna,
     I-40126 Bologna, Italy
\item[\tata] Tata Institute of Fundamental Research, Mumbai (Bombay) 400 005, India
\item[\ne] Northeastern University, Boston, MA 02115, USA
\item[\bucharest] Institute of Atomic Physics and University of Bucharest,
     R-76900 Bucharest, Romania
\item[\budapest] Central Research Institute for Physics of the
     Hungarian Academy of Sciences, H-1525 Budapest 114, Hungary$^{\ddag}$
\item[\mit] Massachusetts Institute of Technology, Cambridge, MA 02139, USA
\item[\panjab] Panjab University, Chandigarh 160 014, India
\item[\debrecen] KLTE-ATOMKI, H-4010 Debrecen, Hungary$^\P$
\item[\dublin] UCD School of Physics, University College Dublin,
 Belfield, Dublin 4, Ireland
\item[\florence] INFN Sezione di Firenze and University of Florence,
     I-50125 Florence, Italy
\item[\cern] European Laboratory for Particle Physics, CERN,
     CH-1211 Geneva 23, Switzerland
\item[\wl] World Laboratory, FBLJA  Project, CH-1211 Geneva 23, Switzerland
\item[\geneva] University of Geneva, CH-1211 Geneva 4, Switzerland
\item[\hamburg] University of Hamburg, D-22761 Hamburg, Germany
\item[\hefei] Chinese University of Science and Technology, USTC,
      Hefei, Anhui 230 029, China$^{\triangle}$
\item[\lausanne] University of Lausanne, CH-1015 Lausanne, Switzerland
\item[\lyon] Institut de Physique Nucl\'eaire de Lyon,
     IN2P3-CNRS,Universit\'e Claude Bernard,
     F-69622 Villeurbanne, France
\item[\madrid] Centro de Investigaciones Energ{\'e}ticas,
     Medioambientales y Tecnol\'ogicas, CIEMAT, E-28040 Madrid,
     Spain${\flat}$
\item[\florida] Florida Institute of Technology, Melbourne, FL 32901, USA
\item[\milan] INFN-Sezione di Milano, I-20133 Milan, Italy
\item[\moscow] Institute of Theoretical and Experimental Physics, ITEP,
     Moscow, Russia
\item[\naples] INFN-Sezione di Napoli and University of Naples,
     I-80125 Naples, Italy
\item[\cyprus] Department of Physics, University of Cyprus,
     Nicosia, Cyprus
\item[\nymegen] Radboud University and NIKHEF,
     NL-6525 ED Nijmegen, The Netherlands
\item[\caltech] California Institute of Technology, Pasadena, CA 91125, USA
\item[\perugia] INFN-Sezione di Perugia and Universit\`a Degli
     Studi di Perugia, I-06100 Perugia, Italy
\item[\peters] Nuclear Physics Institute, St. Petersburg, Russia
\item[\cmu] Carnegie Mellon University, Pittsburgh, PA 15213, USA
\item[\potenza] INFN-Sezione di Napoli and University of Potenza,
     I-85100 Potenza, Italy
\item[\prince] Princeton University, Princeton, NJ 08544, USA
\item[\riverside] University of Californa, Riverside, CA 92521, USA
\item[\rome] INFN-Sezione di Roma and University of Rome, ``La Sapienza",
     I-00185 Rome, Italy
\item[\salerno] University and INFN, Salerno, I-84100 Salerno, Italy
\item[\ucsd] University of California, San Diego, CA 92093, USA
\item[\sofia] Bulgarian Academy of Sciences, Central Lab.~of
     Mechatronics and Instrumentation, BU-1113 Sofia, Bulgaria
\item[\korea]  The Center for High Energy Physics,
     Kyungpook National University, 702-701 Taegu, Republic of Korea
\item[\taiwan] National Central University, Chung-Li, Taiwan, China
\item[\tsinghua] Department of Physics, National Tsing Hua University,
      Taiwan, China
\item[\purdue] Purdue University, West Lafayette, IN 47907, USA
\item[\psinst] Paul Scherrer Institut, PSI, CH-5232 Villigen, Switzerland
\item[\zeuthen] DESY, D-15738 Zeuthen, Germany
\item[\eth] Eidgen\"ossische Technische Hochschule, ETH Z\"urich,
     CH-8093 Z\"urich, Switzerland
\item[\S]  Supported by the German Bundesministerium
        f\"ur Bildung, Wissenschaft, Forschung und Technologie.
\item[\ddag] Supported by the Hungarian OTKA fund under contract
numbers T019181, F023259 and T037350.
\item[\P] Also supported by the Hungarian OTKA fund under contract
  number T026178.
\item[$\flat$] Supported also by the Comisi\'on Interministerial de Ciencia y
        Tecnolog{\'\i}a.
\item[$\sharp$] Also supported by CONICET and Universidad Nacional de La Plata,
        CC 67, 1900 La Plata, Argentina.
\item[$\triangle$] Supported by the National Natural Science
  Foundation of China.
\item[$\clubsuit$] {Visitor from MTA KFKI RMKI, H-1525 Budapest 114, Hungary,
   and Dept.\ of Physics, Harvard University, 17 Oxford St., Cambridge, MA 02138, U.S.A.,
   sponsored by the Scientific Exchange between Hungary (OTKA) and The Netherlands (NWO),
   project B64-27/N25186;  also supported by Hungarian OTKA grants T49466 and NK73143 and
   a HAESF Senior Leaders and Scholars Fellowship.}
\item[$\diamondsuit$] {Present address: The King's University College, Edmonton, Alberta T6B 2H3, Canada}
\item[$\spadesuit$] {Present address: Dept.\ of Business Mathematics and Informatics, K\'aroly R\'obert
   College, H-3200 Gy\"ongy\"os, Hungary}
\item[$\dagger$] Deceased.
\end{list}
}
\vfill
 

\newpage
 
\bibliographystyle{\lthreebiblio/l3style}
\bibliography{%
l3,%
generators,%
jets,%
\lthreebiblio/l3pubs,%
\lthreebiblio/aleph,%
\lthreebiblio/delphi,%
\lthreebiblio/opal,%
\lthreebiblio/markii,%
\lthreebiblio/otherstuff,%
mult,%
bec,levy%
}

\end{document}